%% file: main.tex
\documentclass[a4paper,fleqn,usenatbib]{mnras}

\usepackage{etoolbox}
\makeatletter
\newcommand\thefontsize{The current font size is: \f@size pt}
\newcommand\thefont{\expandafter\string\the\font}
\makeatother

\usepackage[T1]{fontenc}
\usepackage{ae,aecompl}

\usepackage{graphicx}	
\usepackage{amsmath}	
\usepackage{amssymb}	
\usepackage{import}
\import{Aux/}{myPackages.tex}

\usepackage{newtxtext,newtxmath}

\import{Aux/}{myCommands.tex}

\import{Aux/}{otherDefinitions.tex}

\newglossary[cod]{codes}{cod}{ntn}{Codes}
\makeglossaries
\loadglsentries{Glossaries/CodesGlossary}
\loadglsentries{Glossaries/AcronymsGlossary}

\title[Parameter inference with non-linear galaxy clustering]{Parameter inference with non-linear galaxy clustering: accounting for theoretical uncertainties}

\author[]{Mischa Knabenhans${}^{1}$\;\href{https://orcid.org/0000-0002-2886-9838}
{\includegraphics[scale=0.75]{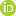}}\;\thanks{E-mail: mischak@physik.uzh.ch},
Thejs Brinckmann ${}^{2,3,4}$\,\href{https://orcid.org/0000-0002-1492-5181}{\includegraphics[scale=0.75]{Img/orcid_16x16.png}}\;\thanks{E-mail: thejs.brinckmann@gmail.com},
Joachim Stadel${}^{1}$\,\href{https://orcid.org/0000-0001-7565-8622}{\includegraphics[scale=0.75]{Img/orcid_16x16.png}},
\newauthor
Aurel Schneider$^{1}$\,\href{https://orcid.org/}{\includegraphics[scale=0.75]{Img/orcid_16x16.png}},
Romain Teyssier$^{1,5,6}$\,\href{https://orcid.org/0000-0001-7689-0933}
{\includegraphics[scale=0.75]{Img/orcid_16x16.png}}
\\
${}^{1}$Institute for Computational Science, University of Zurich, Winterthurerstrasse 190, 8057 Zurich, Switzerland\\
${}^{2}$Dipartimento di Fisica e Scienze della Terra, Universit\'a degli Studi di Ferrara, via Giuseppe Saragat 1, 44122 Ferrara, Italy\\
${}^{3}$Istituto Nazionale di Fisica Nucleare (INFN), Sezione di Ferrara, Via Giuseppe Saragat 1, 44122 Ferrara, Italy\\
${}^{4}$C.N. Yang Institute for Theoretical Physics and Department of Physics \& Astronomy, Stony Brook University, Stony Brook, NY 11794, USA\\
${}^{5}$ Department of Astrophysical Sciences, Princeton University, 4 Ivy Lane, Princeton, NJ 08544, USA\\
${}^{6}$ The Program in Applied and Computational Mathematics, Princeton University, Fine Hall, Washington Road, Princeton, NJ 08544-1000 USA}

\date{Accepted XXX. Received YYY; in original form ZZZ}

\pubyear{2021}

\begin{document}
\label{firstpage}
\maketitle

\begin{abstract}
We implement EuclidEmulator (version 1), an emulator for the non-linear correction of the matter power spectrum, into the \gls{MCMC} forecasting code MontePython. We compare the performance of {\Halofit}, {\HMCode}, and {\EEone}, both at the level of power spectrum prediction and at the level of posterior probability distributions of the cosmological parameters, for different cosmological models and different galaxy power spectrum wave number cut-offs. We confirm that the choice of the power spectrum predictor has a non-negligible effect on the computed sensitivities when doing cosmological parameter forecasting, even for a conservative wave number cut-off of $0.2\hompc$. We find that {\EEone} is on average up to $17\%$ more sensitive to the cosmological parameters than the other two codes, with the most significant improvements being for the Hubble parameter of up to $42\%$ and the equation of state of dark energy of up to $26\%$, depending on the case. In addition, we point out that the choice of the power spectrum predictor contributes to the risk of computing a significantly biased mean cosmology when doing parameter estimations. For the four tested scenarios we find biases, averaged over the cosmological parameters, of between 0.5 and 2$\sigma$ (from below $1\sigma$ up to $6\sigma$ for individual parameters). This paper provides a proof of concept that this risk can be mitigated by taking a well-tailored theoretical uncertainty into account as this allows to reduce the bias by a factor of 2 to 5, depending on the case under consideration, while keeping posterior credibility contours small: the standard deviations are amplified by a factor of $\leq1.4$ in all cases.
\end{abstract}



\import{Chapters/}{01introduction}
\import{Chapters/}{02theory}
\import{Chapters/}{03MCMCs}
\import{Chapters/}{04Forecasts}
\import{Chapters/}{05Estimation}
\import{Chapters/}{06Conclusion}


\section*{Glossary}
\label{glossary}
\glsfindwidesttoplevelname
\setglossarystyle{alttree}
\printglossary[type=main,title=Codes:]
\printglossary[type=acronym,title=Acronyms:]
\nopagebreak

\section*{Acknowledgements}
\label{ackns}
MK acknowledges support from the Swiss National Science Foundation (SNF) grant 200020\_149848. Simulations were performed on the zBox4+ cluster at the University of Zurich. TB was supported by the grant DOE DE-SC0017848, through the INFN project "GRANT73/Tec-Nu", and from the COSMOS network (www.cosmosnet.it) through the ASI (Italian Space Agency) Grants 2016-24-H.0 and 2016-24-H.1-2018. Further, the authors are very grateful for Tim Sprenger's helpful discussions and inputs.

\section*{Data availability}
\label{data_availability}
The data underlying this article will be shared on reasonable request to the corresponding authors.




\bibliographystyle{mnras}
\bibliography{References/gcfc_references.bib}



\appendix
\import{Apps/}{A_ImplementationDetails}

\import{Apps/}{B_MCMCresults}

\bsp	
\label{lastpage}
\end{document}

%% file: Aux/myPackages.tex
\usepackage{euclid}
\usepackage[section]{placeins}
\usepackage{float}
\floatstyle{plaintop}
\usepackage[tableposition=top]{caption}
\restylefloat{table}
\usepackage{comment}
\usepackage{tablefootnote}
\usepackage[acronyms,nogroupskip,nopostdot]{glossaries}
\usepackage{glossary-mcols}
\usepackage{appendix}
\usepackage{xcolor}
\usepackage[colorinlistoftodos]{todonotes}
\usepackage{enumitem}
\usepackage{float}
\usepackage{multirow}
\usepackage{ulem}
\usepackage{cleveref}
\usepackage{dblfloatfix}

%% file: Aux/myCommands.tex

\newcommand{\BAO}{\gls{BAO}}
\newcommand{\BAOs}{\glspl{BAO}}

\newcommand{\CMB}{\gls{CMB}}

\newcommand{\DE}{\gls{DE}}

\newcommand{\EoS}{\gls{EoS}}

\newcommand{\LCDM}{$\Lambda$CDM}

\newcommand{\MCMC}{\gls{MCMC}}
\newcommand{\MCMCs}{\glspl{MCMC}}

\newcommand{\NLC}{\gls{NLC}}

\newcommand{\PT}{\gls{PT}}

\newcommand{\RSD}{\gls{RSD}}
\newcommand{\RSDs}{\glspl{RSD}}

\newcommand{\wCDM}{$w_0$CDM}

\newcommand{\omm}{$\omega_{\mathrm m}$}

\newcommand{\h}{$h$}

\newcommand{\CLASS}{\gls{code:CLASS}}

\newcommand{\EEone}{\gls{code:EuclidEmulator1}}
\newcommand{\EEtwo}{\gls{code:EuclidEmulator2}}

\newcommand{\Halofit}{\gls{code:HALOFIT}}
\newcommand{\HMCode}{\gls{code:HMCode}}

\newcommand{\MontePython}{\gls{code:MontePython}}

\newcommand{\ve}{\mathbf{e}}
\newcommand{\vk}{\mathbf{k}}

\newcommand{\hompc}{\,h\,{\rm Mpc}^{-1}}




%% file: Aux/otherDefinitions.tex
\definecolor{mygreen}{RGB}{26,120,34}
\definecolor{myred}{RGB}{200,0,33}
\definecolor{myorange}{RGB}{233,133,46}
\definecolor{mypurple}{RGB}{120,0,100}
\definecolor{mycyan}{RGB}{56,200,252}
\definecolor{mymagenta}{RGB}{208,102,142}
\definecolor{myyellow}{RGB}{254,197,52}

\newlist{inparaenum}{enumerate}{2}
\setlist[inparaenum]{nosep}
\setlist[inparaenum,1]{label=\bfseries\arabic*.}
\setlist[inparaenum,2]{label=\arabic{inparaenumi}\emph{\alph*})}

%% file: Glossaries/CodesGlossary.tex
\newglossaryentry{code:CAMB}
{
  name=\texttt{CAMB},
  description={Code for anisotropies in the microwave background},
}
\newglossaryentry{code:CLASS}
{
  name=\texttt{CLASS},
  description={Cosmological linear anisotropy solving system},
}

\newglossaryentry{code:classy}
{
  name=\texttt{classy},
  description={Python wrapper for CLASS},
}

\newglossaryentry{code:CONCEPT}
{
  name=\texttt{CO\textit{N}CEPT},
  description={Cosmological N-body code written in Python},
}

\newglossaryentry{code:CosmicEmu}
{
  name=\texttt{CosmicEmu},
  description={Cosmic emulator based on the Mira-Titan cosmological simulation suite (successor of FrankenEmu based on the Coyote simulation suite).},
}
\newglossaryentry{code:EuclidEmulator1}
{
  name=\texttt{EuclidEmulator1},
  description={Emulator code to emulate non-linear corrections (boost factors) to DM power spectra},
}

\newglossaryentry{code:EuclidEmulator2}
{
  name=\texttt{EuclidEmulator2},
  description={Successor of EuclidEmulator1. Emulator code to emulate non-linear corrections (boost factors) to DM power spectra self-consistently including corrections from massive neutrinos and dark energy perturbations.},
}

\newglossaryentry{code:e2py}
{
  name=\texttt{e2py},
  description={Python wrapper for EuclidEmulator},
}

\newglossaryentry{code:HACC}
{
  name=\texttt{HACC},
  description={Hardware/Hybrid Accelerated Cosmology Code},
}

\newglossaryentry{code:HALOFIT}
{
  name=\texttt{HALOFIT},
  description={Analytical code to produce non-linear power spectra},
}

\newglossaryentry{code:HMCode}
{
  name=\texttt{HMCode},
  description={Analytical code to produce non-linear power spectra},
}

\newglossaryentry{code:MontePython}
{
  name=\texttt{MontePython},
  description={Monte Carlo code for Cosmological Parameter extraction. It contains likelihood codes of most recent experiments, and interfaces with the Boltzmann code class for computing the cosmological observables.},
}
\newglossaryentry{code:NGenHalofit}
{
  name=\texttt{NGenHalofit},
  description={Code to produce non-linear power spectra using a semi-analytical approach for large and a smoothing-spline-fit model for small scales},
}

\newglossaryentry{code:PKDGRAV3}
{
  name=\texttt{PKDGRAV3},
  description={parallel k-D tree gravity code (version 3); Cosmological N-body tree code},
}
\newglossaryentry{code:UQLab}
{
  name=\texttt{UQLab},
  description={Matlab-based uncertainty quantification framework},
}

%% file: Glossaries/AcronymsGlossary.tex
\newacronym{BAO}{BAO}{Baryon accoustic oscillations}
\newacronym{CMB}{CMB}{cosmic microwave background}
\newacronym{DE}{DE}{Dark energy}
\newacronym{DM}{DM}{Dark matter}
\newacronym{ED}{ED}{experimental design}
\newacronym{EE}{EE}{EuclidEmulator}
\newacronym{EFHR}{EFHR}{Euclid Flagship High Resolution}
\newacronym{EOE}{EOE}{emulation-only error}
\newacronym{EoS}{EoS}{equation of state}
\newacronym{GPs}{GPs}{Gaussian processes}
\newacronym{GR}{GR}{general theory of relativity}
\newacronym{ICs}{ICs}{initial conditions}
\newacronym{LARS}{LARS}{Least angle regression-based selection}
\newacronym{LH}{LH}{Latin hypercube}
\newacronym{LHS}{LHS}{Latin hypercube sampling}
\newacronym{LPT}{LPT}{1st order Lagrangian perturbation theory}
\newacronym{LV}{LV}{Large volume}
\newacronym{HOD}{HOD}{halo occupation distribution}
\newacronym{MCMC}{MCMC}{Markov chain Monte Carlo}
\newacronym{NLC}{NLC}{nonlinear correction}
\newacronym{PC}{PC}{principal component}
\newacronym{PCA}{PCA}{principal component analysis}
\newacronym{PCE}{PCE}{polynomial chaos expansion}
\newacronym{PCS}{PCS}{piece-wise cubic spline}
\newacronym{PT}{PT}{perturbation theory}
\newacronym{QNL}{QNL}{quasi-non-linear}
\newacronym{RSD}{RSD}{redshift space distortion}
\newacronym{PM}{PM}{particle-mesh}
\newacronym{PPF}{PPF}{parametrized post-Friedmann}
\newacronym{SPCE}{SPCE}{sparse polynomial chaos expansion}
\newacronym{THF}{THF}{Takahashi HALOFIT}
\newacronym{UQ}{UQ}{uncertainty quantification}
\newacronym{ZA}{ZA}{Zel'dovich approximation}

%% file: Chapters/01introduction.tex
\section{Introduction}
\label{sec:introduction}
After very successful missions such as Planck and SDSS, many ground and space based experiments are being conducted or planned in order to learn more about our Universe. The Dark Energy Spectroscopic Instrument\footnote{www.desi.lbl.gov/category/announcements/} \citep[DESI,][]{DESICollaboration2016}, the European space mission Euclid\footnote{sci.esa.int/euclid} \citep{Laureijs2011EuclidReport}, the Legacy Survey of Space and Time (LSST) of the Vera C. Rubin observatory\footnote{https://www.vro.org, www.lsst.org/lsst} \citep{LSSTScienceCollaboration2009}, the Nancy Grace Roman Space Telescope\footnote{http://roman.gsfc.nasa.gov} \citep{Akeson2019}, and the Square-Kilometre Array\footnote{https://www.skatelescope.org} \citep[SKA,][]{Maartens2015_SKA} are a few well-known examples. They will provide highly accurate measurements of various cosmological observables, giving insight into the nature of dark matter and dark energy, and are likely to measure the total mass of neutrinos.

In order to exploit this wealth of high-quality data that is being produced, it has to be met by equally accurate theoretical predictions. We need to quantify the process of structure formation at high precision and to include all systematics related to the observations. Only if the highly demanding requirements on both sides are met, do we optimize our chances to determine the cosmological parameters with high enough accuracy to potentially reveal new physics. 

One approach is extended perturbation theories that are able to accurately model the mildly non-linear scales. See e.g. CLASS-PT~\citep{Chudaykin2020}, which was recently used for Euclid forecasts~\citep{Chudaykin2019} and BOSS data analyses~\citep{Ivanov2020CosmologicalParametersFromBoss, Ivanov2020CosmoParamsfromPlanckBoss}, and EFTofLSS (publicly available as PyBird,~\citealt{DAmico2020PyBird}), also recently used for BOSS data analyses~\citep{Colas2020,DAmico2020LSSEFT}. These approaches allow us to perturbatively compute an accurate galaxy power spectrum to a wave number of about $k = 0.3 \hompc$, exceeding the range of reliability of linear perturbation theory.

However, future surveys are expected to probe much smaller scales, e.g. Euclid will collect accurate data up to $k = 10 \hompc$. In order to accurately model increasingly non-linear scales, those methods either need drastic improvements, or we need to use a different approach, e.g. other non-linear modelling techniques such as emulation. Cosmic emulators do not suffer from the neuralgic problems at high $k$ values as do perturbation theoretical approaches, nor do they require expensive N-body simulations beyond those used initially to train the emulator. Such emulators have received significant attention in recent years with the development of new publicly available tools and codes \citep{Heitmann2009, Heitmann2010, Lawrence2010, Heitmann2013, Heitmann2016, Mead2015, Mead2016, Lawrence2017, DeRose2019, McClintock2019TheFunction, Zhai2018, Nishimichi2019, Rogers2019, Valcin2019, Bird2019AnForest, Winther2019, McClintock2019, Fluri2019CosmologicalMaps, Cataneo2019OnGravity, Angulo2020, Mead2021}.

We present an improvement on prediction accuracy of galaxy clustering observables thanks to integration of the publicly available {\EEone}\footnote{We use version 1 of \texttt{EuclidEmulator}, referred to as \EEone~, which is available at \url{https://github.com/miknab/EuclidEmulator}. Note that during the preparation of this manuscript an improved version, \EEtwo~\citep{Knabenhans2021}, has been made publicly available at \url{https://github.com/miknab/EuclidEmulator2}. The latter features additional cosmological parameters, namely the neutrino mass sum ($\sum m_\nu$) and dynamical dark energy ($w_0$,$w_a$), as well as an expanded redshift and wave number range, and an improved python wrapper.}\citep{Knabenhans2019} into the {\MCMC} sampling package MontePython\footnote{See \url{https://github.com/brinckmann/montepython_public} for the latest version 3.5.} \citep{Audren2013ConservativePYTHON, Brinckmann2019}. {\EEone} is a cosmic emulator that quickly and accurately predicts the non-linear correction factor of the matter power spectrum for cold dark matter cosmologies with a varying time-independent {\DE} {\EoS} parameter $w_0$ (these are commonly referred to as ``$w_0$CDM'' cosmologies).

Popular methods for including non-linear information of cosmic structure formation in an {\MCMC} run are via {\Halofit} \citep{Smith2003, Takahashi2012} or {\HMCode}\footnote{Since {\HMCode}-2020 ~\citep{Mead2021} is not yet available within the Boltzmann solver CLASS, we instead make use of {\HMCode}-2016. Note that the older version is known to have difficulties in capturing non-linear features correctly, e.g. non-linear damping of the {\BAO}, which were improved in the newer version. While this may impact some parts of the analysis, it also illustrates the need for accurately accounting for theoretical uncertainties and problems associated with not doing so.\label{footnote:hmcode}}~\citep{Mead2015, Mead2016}. In this work we compare these two codes to a cosmic emulator, {\EEone}, which is able to capture non-linear features, e.g. non-linear damping of the {\BAO}, as accurately as $0.3\%$ or less compared to full N-body simulations over a wide range of both redshifts and $k$ modes -- an accuracy level unprecedented by any other cosmic emulator.

The most widely used forecasting method is Fisher matrix forecasts (e.g. \citealt{Tegmark1997Karhunen-LoeveSets}). In its traditional form, it is conceptually very simple, but is restricted to certain very idealistic mathematical properties of the corresponding likelihood function (efforts in development of methods trying to overcome this limitation can, e.g., be found in \citealt{Sellentin2014BreakingMatrices}). Further, it involves computations of numerical derivatives that are likely to be unstable. For a discussion of these problems and efforts towards solutions we refer to \citet{Blanchard2019EuclidISTForecast, YahiaCherif2020, Bhandari2021_arXiv}. 

We connect measurements and theory in a Bayesian forecasting process through {\MCMC} sampling method \citep{Christensen2001BayesianMeasurements, Lewis2002,Perotto2006ProbingSimulations}, that estimates the posterior probability distributions of cosmological parameters. This is a more data driven, iterative approach, that has the advantage of not relying on derivatives nor on any assumptions of the shape of the underlying likelihood. This is the most common approach for cosmological parameter inference. Since it provides more reliable results in parameter sensitivity forecasts than classical Fisher analyses, we choose to use the {\MCMC} method in the analysis presented in this work.

In this work, we investigate the influence of different code implementations (with different strength and weaknesses) on sensitivity forecast and parameter estimation results and we argue that using theoretical uncertainties can help deal with that (also beyond forecasts in which they are used already today). We choose to perform this analysis with the following three codes: {\Halofit} and {\HMCode} because they are very established in the field and {\EEone} as it is a representative of cosmic emulators which, as described above, have become more popular recently . We shall stress, however, that the key take-away message of this work is of qualitative rather than quantitative nature and that these qualitative results are ultimately independent of the set of codes used. Along the way of our investigations we address four main questions:
\begin{enumerate}
    \item What are the performance differences of {\EEone}, {\Halofit} and {\HMCode}?
    \item What is the added value of considering (mildly) nonlinear scales in parameter forecasts?
    \item What is the impact of the choice of the (non-linear) predictor model on the parameter estimation result?
    \item How are the forecasting results affected by different choices of theoretical uncertainty models?
\end{enumerate}
In order to understand (i), we first compare the codes at the level of the matter power spectrum in \autoref{sec:code_agreement} (see \autoref{fig:zvar} and ~\ref{fig:var}) before performing a thorough forecast analysis in \autoref{sec:forecasts}. 

Traditionally, in galaxy clustering analyses information from non-linear scales is removed as being "too difficult" to model correctly, and usually only information from linear and quasi-linear scales is included. However, a wealth of information is available on non-linear scales that may improve constraints. Indeed, a few parameters exhibit very clear signals only in the (mildly) non-linear regime, particularly the {\DE} {\EoS} parameter $w_0$, which is illustrated in \autoref{fig:var}. As such, we address question (ii) in \autoref{subsec:kmax_uncertainties} by increasing the maximum wavenumber included in the analysis. This makes a case for cosmic emulators as the accuracy at nonlinear scales is one of their key selling points in contrast to models based on {\PT}.

However, once we go beyond linear scales, theoretical modelling uncertainties grow substantially, as different simulation codes increasingly disagree with each other. While the gravity-only calculation is under control up to high $k$-modes \citep{Schneider2015,Garrison2019}, baryonic feedback effects are known to affect the matter power spectrum beyond $k\sim 0.1-1$ h/Mpc but remain poorly constrained \cite[see e.g.][]{VanDaalen2011,Schneider2015,Chisari2019,Arico2020,Mead2021,Giri2021}.

\citet{Sprenger2019} introduced a theoretical uncertainty increasing with wavenumber in order to account for the growing uncertainties at small cosmological scales, thereby making a step in the direction towards including more non-linear $k$ modes into the analysis rather than just ignoring them as was often done in the past (see~\citealt{Audren2013NeutrinoErrors,Baldauf2016} for earlier work in this direction). An uncertainty like that used in \citet{Sprenger2019} works well for sensitivity forecasts, where we are only interested in the uncertainties of parameter estimates and not the estimates themselves. However, results presented in this paper indicate that in the case of (mock) data analysis parameter estimation we should consider a theoretical uncertainty directly tied to the modelling choices made for the specific code that is used to fit the data, as otherwise the parameter estimates are likely to be biased.

While we discuss theoretical uncertainties and their modelling in \autoref{sec:envelope}, they are also the focus of questions (iii) and (iv), which are addressed in \autoref{subsec:code_dependence} and \ref{subsec:tue}. Indeed, when we set up a scenario to mimic real data analysis (where the true model is not known), we find large biases in the inferred parameters, in particular when extending the analysis to larger wavenumbers. This is not unexpected, as, indeed, the three codes model non-linear scales differently. It serves as a word of caution, that when high resolution data from future surveys such as Euclid is analysed, we need to carefully account for modelling uncertainties. We make a first attempt at this using an inflated theoretical uncertainty inspired by the agreement between {\EEone} and the other two codes, and we show that with such an uncertainty envelope, we get greatly improved agreement between the codes, reducing the bias in the parameter inference without significantly decreasing the sensitivity. We leave it to future work to produce a more robust scheme for accounting for the theoretical modelling uncertainty of current non-linear parameter estimation tools.

This paper is structured as follows: In \autoref{sec:theory} present a comparison of the three codes analyzed in this paper -- {\EEone}, {\Halofit}, and {\HMCode} -- at the level of the matter power spectrum and discuss both known and unknown systematics impacting parameter estimations and sensitivity forecasts as well as how they are related to code differences. In this section we motivate the need for a theoretical uncertainty in order to mitigate unwanted effects due to those systematics. Next, in \autoref{sec:analysis} we introduce the analysis methodology employed in this work, including some theoretical considerations as well as the strategy of our computational experiments. The results are presented and discussed in \autoref{sec:forecasts} and \ref{sec:parameterestimation}. We finally conclude this paper in \autoref{sec:conclusion}.

%% file: Chapters/02theory.tex
\section{Systematic unknowns in non-linear cosmological structure formation}
\label{sec:theory}
\subsection{Code comparison at the matter power spectrum level}
\label{sec:code_agreement}
Before jumping to the \gls{MCMC} results, let us first quantify differences between the three non-linear prescriptions for predicting the matter power spectrum $P(k,z; p)$ considered in this paper: {\Halofit} (``HF''), {\HMCode} (``HM'') and {\EEone} (``EE''). This will help us understand the differences we may see in the following section.

Due to differences in how the non-linear prescriptions model nonlinear structure formation, we expect a difference in how each code responds to a changes in cosmological parameters. In the following, we want to compare the non-linear prescriptions to each other in order to quantify these differences. Taking {\EEone} as the reference, we vary one cosmological parameter at a time, keeping all others fixed to a fiducial cosmological model (Euclid Reference Cosmology as defined in \citealt{Knabenhans2019}). We define the model ratio
\begin{equation} 
R_{\mathrm{model}}(k,z; p) \equiv \frac{P_{\mathrm{model}}(k,z; p)}{P_{\mathrm{EE}}(k,z; p)}
\label{eq:Rmodel}
\end{equation}
where $\mathrm{model}\in\{\mathrm{HF}, \mathrm{HM}\}$ and $p\in\{\omega_{\mathrm{b}}, \omega_{\mathrm{m}}, n_{\mathrm{s}}, h, w_0, A_{\mathrm{s}}\}$\footnote{{\EEone} does not accept $A_\mathrm{s}$ as input but only $\sigma_8$. The conversion for each set of cosmological parameters is done with {\CLASS}.}. We consider the difference with respect to the reference cosmological model
\begin{equation} 
\Delta R_{\mathrm{model}}(k,z; p) = R_{\mathrm{model}}(k,z; p) - R_{\mathrm{model}}(k,z; p_{\mathrm{ref}})
\label{eq:DeltaR}
\end{equation}
This quantifies how, for a given non-linear prescription, the matter power spectrum changes when we vary one of the cosmological parameters.

\begin{figure}
  \centering
  \includegraphics[width=\columnwidth]{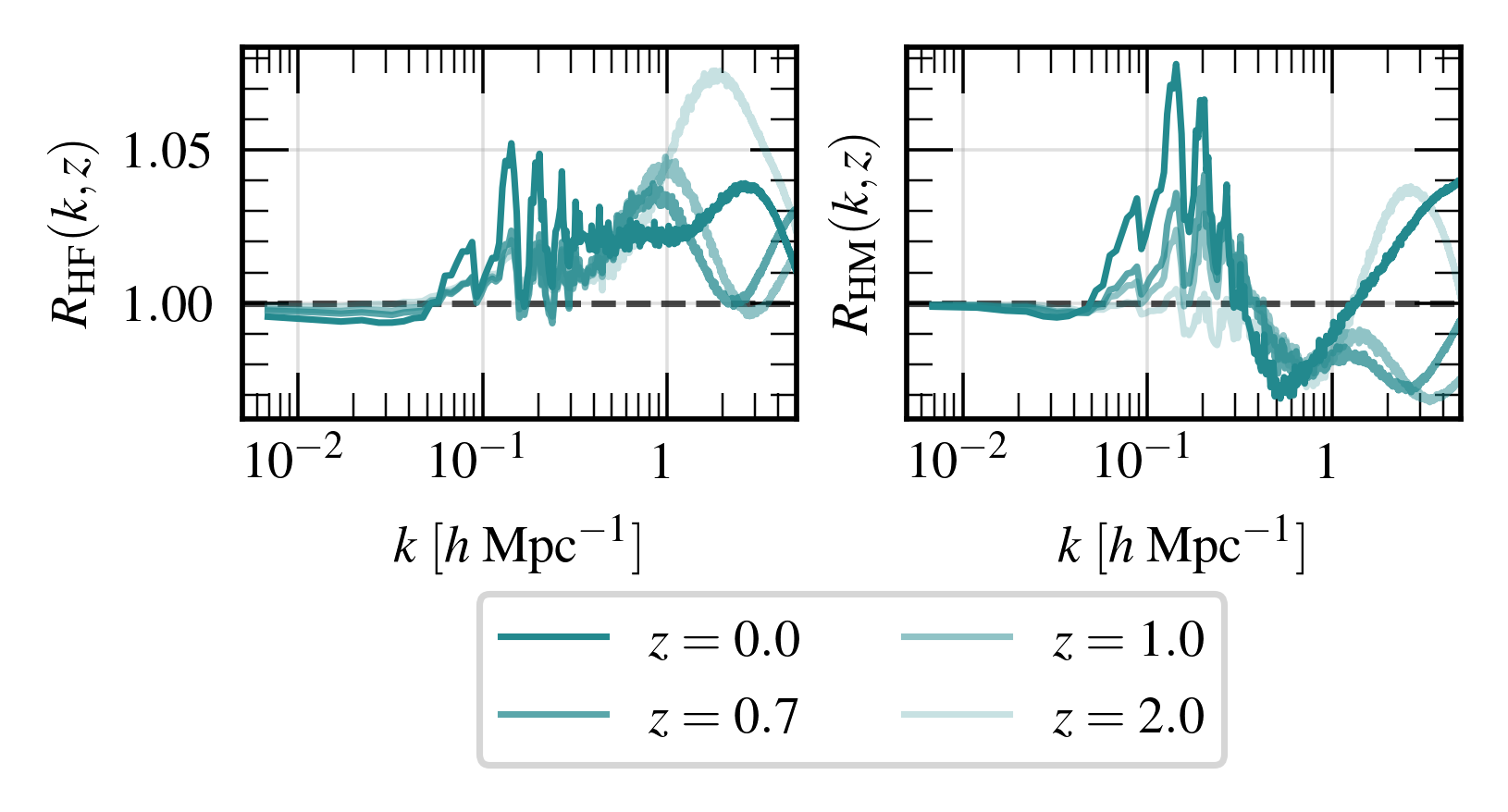}
  \caption{Redshift evolution for power spectrum ratio (as defined in \autoref{eq:Rmodel}) between {\Halofit} and {\EEone} (left panel) and {\HMCode} and {\EEone} (right panel). The evolution, which is only weakly dependent on the cosmology, is shown for the {\Euclid} reference cosmology.}
  \label{fig:zvar}
\end{figure}
    
\begin{figure*}
  \centering
  \includegraphics[width=\textwidth]{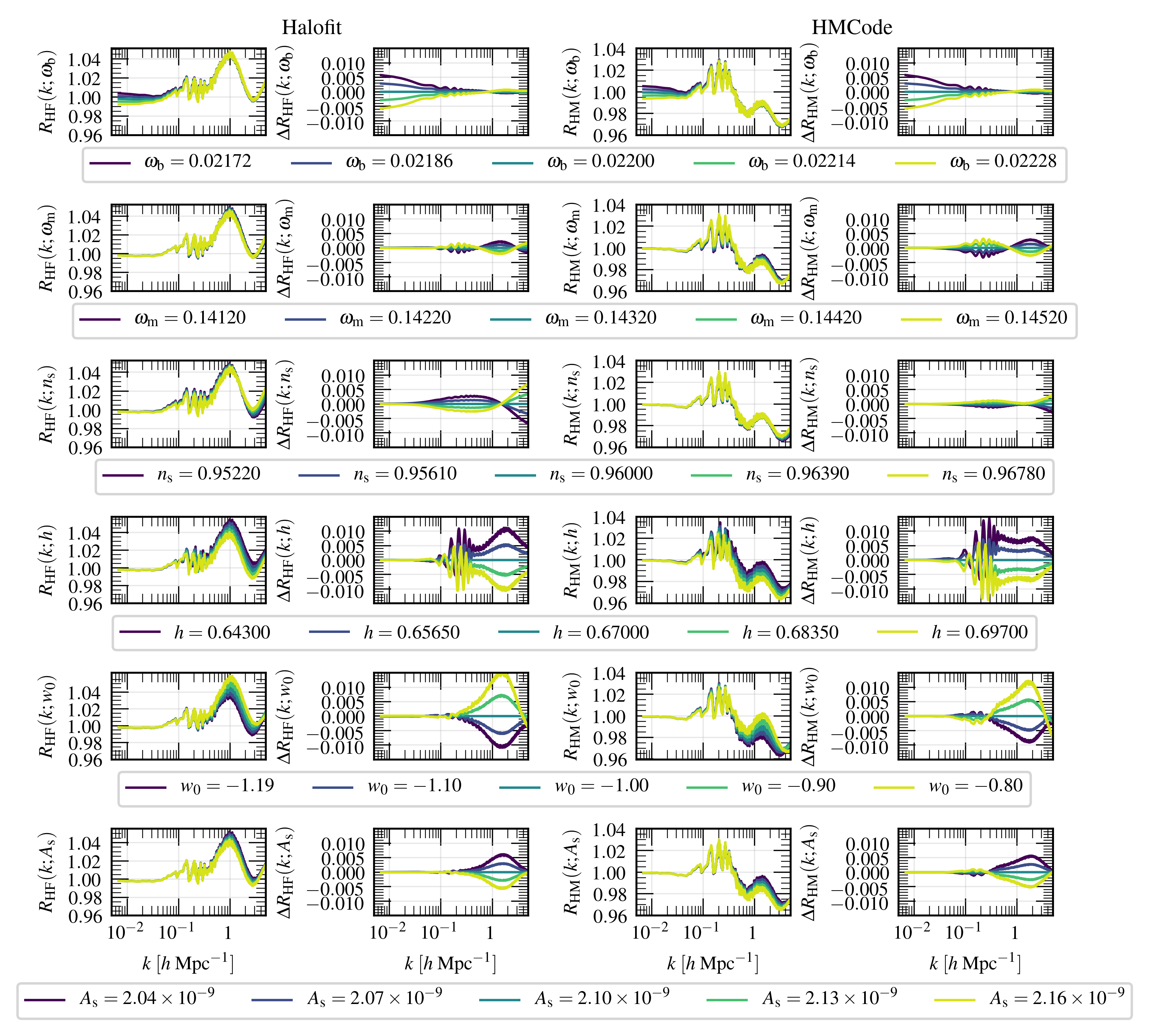}
  \caption{Variation of the non-linear matter power spectrum due to variation of cosmological parameters. Each row shows the variation of one cosmological parameter: $\omega_{\rm{b}}$ in the first row, $\omega_{\rm{m}}$ in the second, etc. All parameters that are not varied in any given row, are set to their corresponding values in the Euclid Reference Cosmology. Notice that the Euclid Reference cosmology corresponds to the middle parameter values in each of the parameter value legends, i.e. $\omega_{\rm{b}} = 0.022$, $\omega_{\rm{m}} = 0.1432$, $n_{\rm{s}} = 0.96$, $h = 0.67$. $A_{\rm s} = 2.1\times10^{-9}$ is always fixed. The variation is shown for the quantity $R_\mathrm{model}$ as defined in \autoref{eq:Rmodel} (columns 1 and 3). In order to emphasize (i) which scales are affected the most by the variation of a specific parameter and (ii) how strongly $R$ is impacted by the variation of each parameter, we also plot the $\Delta R$ with respect to the $R$ of the Euclid Reference Cosmology (columns 2 and 4). It is evident, for instance, that the variation of $h$ and $w_0$ has a much larger effect on $R$ than e.g. $\omega_{\rm{m}}$. Notice that the tested parameter values are taken to be the Euclid Reference Cosmology value (the central value) $\pm i\sigma_\mathrm{p, Planck2018}, i\in{1,2}$ where $\sigma_\mathrm{p, Planck2018}$ denotes the standard deviation of each parameter $p$ of the Planck 2018 results (for more details see section 17.14 on p. 301 in \citealt{PlanckParameterTables}).}
  \label{fig:var}
\end{figure*}

In \autoref{fig:zvar}, we show the ratio from~\autoref{eq:Rmodel}, i.e. how well {\Halofit} (left panel) and {\HMCode} (right panel) agree with {\EEone}. This is shown at four redshifts ($z=[0,0.7,1,2]$), which includes the ratio today and spans the redshift range of the Euclid galaxy clustering likelihood ($z=0.7$ to $z=2$), and for a fixed cosmology (this is also shown for a varying cosmology at $z=1$ in \autoref{fig:var} columns 1 and 3). We show only the redshift dependence for the reference cosmological model, as the difference from~\autoref{eq:DeltaR} changes negligibly across redshift. We see several notable trends:
\begin{itemize}
    \item Both {\Halofit} and {\HMCode} see an overall small shift of the matter power spectrum to smaller scales (larger $k$), resulting in a disagreement on BAO scales ($k \approx [0.05,0.5] \hompc$) as seen from the sharp wiggles and the small overall amplitude shift on those scales. For both codes, the disagreement on BAO scales compared to {\EEone} increases with smaller redshift. The disagreement on BAO scales compared to {\EEone} is obviously highly scale-dependent due to the BAO feature, but it is maximally about $5\%$ for {\Halofit} at $z=0$ (about $2\%$ in the Euclid likelihood redshift range) and about $8\%$ for {\HMCode} at $z=0$ (about $4\%$ in the Euclid likelihood redshift range). For {\Halofit}, we see a larger amplitude of the matter power spectrum compared to {\EEone} across all redshifts considered, while for {\HMCode} the sign of the amplitude shift depends on redshift: at $z=2$ the amplitude is shifted to smaller values, while at the other redshifts considered the amplitude of the matter power spectrum is shifted to larger values. 
    \item On small scales ($k \gtrsim 0.5 \hompc$), we see a very different behavior between each of the three codes. Across all redshifts, {\Halofit} sees a larger amplitude of the matter power spectrum compared to {\EEone}, but in a scale-dependent and redshift-dependent way: at $z=0$ the disagreement compared to {\EEone} peaks at $4\%$ at $k \approx 2.5 \hompc$ and is about $2-3\%$ across the rest of the interval ($k \approx [0.5,5] \hompc$); at $z=0.7$ the disagreement peaks at about $3.5\%$ at $k \approx 0.8 \hompc$, dropping to no difference at $k \approx 2 \hompc$, before rising to $3\%$ at $k \approx 5 \hompc$; at $z=1$ the disagreement peaks at about $4.5\%$ at $k \approx 1 \hompc$, dropping to no difference at $k \approx 3 \hompc$, before rising to $1.5\%$ at $k \approx 5 \hompc$; at $z=2$ the disagreement peaks at about $7.5\%$ at $k \approx 2 \hompc$, dropping to about $2-3\%$ near either end point of the interval $k \approx [0.5,5] \hompc$.
    \item On small scales ($k \gtrsim 0.5 \hompc$) for {\HMCode}, we see a decrease in the amplitude of the matter power spectrum compared to {\EEone} for all redshifts of about 1-3\%, except for $z=0$ and $z=2$, where the disagreement changes from a $3\%$ decrease in amplitude at around $k \approx [0.5,1] \hompc$ to a $2-4\%$ increase at around $k \approx [2,5] \hompc$: at $z=2$ this comes as a peak at $4\%$ at $k \approx 2.5 \hompc$ that again decreases to around $1\%$ at $k \approx 5 \hompc$, while at $z=0$ this comes as a steady increase until a peak at $4\%$ at $k \approx 5 \hompc$.
\end{itemize}

In~\autoref{fig:var}, we vary the cosmological parameters one at a time at redshift $z=1$: the baryon abundance $\omega_{\rm{b}}$ (first row), the matter abundance $\omega_{\rm{m}}$ (second row), the tilt of the primordial power spectrum $n_{\rm{s}}$ (third row), the Hubble parameter $h$ (fourth row), the equation of state of dark energy $w_0$ (fifth row), and the amplitude of the primordial power spectrum $A_{\rm{s}}$. We show the ratio from~\autoref{eq:Rmodel} in columns 1 ({\Halofit}) and 3 ({\HMCode}), and the difference from~\autoref{eq:DeltaR} in columns 2 ({\Halofit}) and 4 ({\HMCode}). Notice that the tested parameter values are taken to be the Euclid Reference Cosmology value (the central value) $\pm i\sigma_\mathrm{p, Planck2018}, i\in{1,2}$ where $\sigma_\mathrm{p, Planck2018}$ denotes the standard deviation of each parameter $p$ of the Planck 2018 results (for more details see section 17.14 on p. 301 in \citealt{PlanckParameterTables})\footnote{Notice that setting $w_0 = w_0^{\rm{ref}}-2\sigma_\mathrm{p, Planck2018} = -1.2$ while keeping all other cosmological parameter fixed to their fiducial values results in a $\sigma_8$ value outside the parameter space accepted by {\EEone}. For this reason the corresponding $w_0$ value was manually adjusted to $w_0 = -1.19$.}.

We see that the difference (\autoref{eq:DeltaR}) of {\Halofit} and {\HMCode} compared to {\EEone} is fairly small for $\omega_{\rm{b}}$, $\omega_{\rm{m}}$, and $n_{\rm{s}}$. For $\omega_{\rm{b}}$, the difference peaks at about $0.5\%$ on very large, unobservable scales (small $k$) for a $2\sigma$ variation in the parameter, with negligible difference between {\Halofit} and {\HMCode}; for $\omega_{\rm{m}}$ the the difference peaks on observable scales, but with only about $0.4\%$ maximal difference for {\HMCode} (the difference is smaller for {\Halofit}) for a $2\sigma$ variation in the parameter; for $n_{\rm{s}}$ the difference is largest for {\Halofit} (it is much smaller for {\HMCode}), peaking at around $0.4-0.6\%$ for a $2\sigma$ variation in the parameter.

For the two parameters, $h$ and $w_0$, we see over percent level differences: for $h$ we see a large difference of {\Halofit} and {\HMCode} compared to {\EEone} across BAO and small scales, with the largest difference on BAO scales, where the shift in BAO peak location results in a difference of up to $1.5\%$ for {\HMCode} and $1.1\%$ for {\Halofit}; for $w_0$ the largest difference is on small scales, ranging from up to a few permille on BAO scales to $1-1.5\%$ (with largest difference compared to {\EEone} being for {\Halofit}) at around $k \approx 1.5-2 \hompc$.

For the last parameter, $A_{\mathrm{s}}$, we see again somewhat smaller differences, that reach their maximal amplitude of about $0.6\%$ at $k\sim1.5 \hompc$ for {\Halofit}. For {\HMCode} the maximal amplitude is marginally smaller (about $0.5\%$) and it is reached only at $k\sim 2\hompc$.

We will see in~\autoref{sec:parameterestimation} that the differences in~\autoref{fig:zvar} lead to parameter estimation biases when performing a mock data analysis. This will be discussed further there.

\subsection{Sensitivity forecasting and parameter estimation at non-linear scales}
\label{subsec:codedisagreementimpact}
In the past, galaxy clustering data analyses used to include a conservative non-linear cut-off in $k$-space, above which all data is discarded. As, however, it is exactly these non-linear scales that modern cosmological surveys aim to exploit, such a cut-off defeats the point of the surveys. Yet, the theory required to analyse the data from non-linear scales is based on imperfect understanding of the non-linear structure evolution in the Universe due to hydrodynamics, galaxy formation, feedback processes etc. As a consequence, the codes implementing the theory (such as {\HMCode}, {\Halofit} and {\EEone} discussed in the previous subsection) tend to increasingly disagree the more non-linear scales are considered.

We will now discuss how such theory and code disagreements can influence cosmological parameter inference on a grander scale. There are two primary tasks/goals in cosmological parameter inference:
\begin{itemize}
    \item Measuring cosmological parameters with high accuracy (low bias)
    \item Measuring cosmological parameters with high precision (low variance/uncertainties) in a \textit{robust} manner
\end{itemize}
For both tasks, we typically compare a given theory to (mock) data by computing (or sampling) a posterior distribution for the cosmological parameters in a Bayesian way: the parameter value is estimated by the maximum posterior probability while the uncertainties are derived from the credible contours of the posterior. As credible contours can be reduced by simply ignoring sources of uncertainty, robustness plays also a key role in precision measurements: we try to find the credible contours that are simultaneously as small and as correct as possible.

``Parameter estimation'' is the task in which the values of the parameters themselves are to be determined, i.e. in which the maximum posterior probability is to be evaluated. In a real-case scenario, the data comes from an actual observation/experiment. In past decades, data was mainly collected from the realm of linear structure formation, which stands on firm theoretical grounds such that there was little disagreement about the theories used to analyze the data.

However, modern cosmological experiments have entered the realm of non-linear structure formation and the corresponding data contains a wealth of valuable information which should not be disregarded. Nevertheless, non-linear structure formation is not known at the same level of precision as its linear counterpart and, hence, it is important to remember that the choice of the predictor may lead to slightly different estimates of the parameter values. In other words, the obtained parameter values are not universal but ``remember'' the theory/code they were estimated with. A similar result has been found and published by \citet{Martinelli2021}, who used weak lensing observables for their analysis. In this paper, we offer a proof-of-concept solution for this problem inspired by work previously published: we suggest that code-dependence is best handled by introducing a ``theoretical uncertainty'' into the data analysis (in \citealt{Audren2013NeutrinoErrors,Baldauf2016, Sprenger2019} this concept is introduced solely in the context of sensitivity forecasting). Theoretical uncertainties are discussed more deeply in \autoref{subsec:tue}.\\
\\
``Parameter sensitivity forecasting'' (or short sensitivity forecasting) is a common method to project how precisely a given experiment is able to determine the cosmological parameters. As such, it provides guidance in decision making processes regarding the experimental design of a given survey and verifies to the community that the science goals are achievable. 

In contrast to the task of parameter estimation, sensitivity forecasts are usually performed with mock data. This mock data is typically generated using the same code and theory as is used to compute/sample the posterior distribution, thereby we assume that the data is perfectly modelled with appropriate uncertainties. Developments in cosmological structure formation theory, and the ever increasing accuracy of computational methods able to predict cosmological observables at non-linear scales, allow scientists to better take into account systematics in their analyses. For this reason, it is an interesting exercise in its own right to measure the performance of newly developed methods applied to sensitivity forecasts and benchmark them against more established tools. However, even when various sources of uncertainties are taken into account, it is highly unlikely that none have been missed at all.

In the context of sensitivity forecasts, theoretical uncertainties can hence be used in order to (1) fold in uncertainties due to neglected systematics (such as e.g. baryonic physics or scale-dependent galaxy bias) and (2) to deal with unknown uncertainties that have not been modelled. As, however, different codes implement different theories and assumptions and hence take into account different sources of uncertainties, the result of sensitivity forecast does also depend on the choice of code used. A theoretical uncertainty hence makes a sensitivity forecast more robust and less dependent on the modelling choices implemented into the codes used in the forecast.

\subsection{Treatment of known systematics: parameter inference with cosmological emulators}
There is a growing number of power spectrum predictors available on the market, ranging from full-fledged cosmological simulations over halo model derivatives and {\PT}-based models all the way to purely data-driven approaches. The first kind offers the most accurate predictions but are not useful in neither have limited use for sensitivity forecasts and parameter estimations due to their immense computational expense. {\PT}- and halo model-based models are orders of magnitudes cheaper to evaluate than simulations but they suffer from more systematics as they are based on calibrated fitting functions. In the past, they were the choice of power spectrum predictors in parameter inference thanks to their low computational cost. However, emulators are comparable with (semi-)analytical models in terms of computational expense but often carry less systematics as they are based on simulations. It must be emphasized, though, that the remaining systematics need to be treated carefully. Observations necessarily contain information whose modelling needs to take into account sources of uncertainty such as {\RSDs}, the Alcock-Paczy\'nski effect, galaxy bias, baryonic effects on structure formation etc. (see~\autoref{sec:galaxypower}).

Although this is not a special trait of emulators, many are actually designed to predict correction factors to deal with individual systematics. Examples are e.g. the non-linear correction of the matter power spectrum \citep{Knabenhans2019, Knabenhans2021}, the baryonic emulator \citep{Schneider2020_BaryonicEffects1,Giri2021}, or the halo bias \citep{Valcin2019, McClintock2019}. This way, multiple individual emulators, each modelling a different effect, can, under certain assumptions, be stacked on top of each other to take a multitude of systematics into account.

It can therefore be anticipated that future parameter inference studies will use emulators as robust predictors that may include uncertainties in a modular way. A possible approach to deal with unknown sources of systematic errors is discussed in the subsequent subsection.

\subsection{Treatment of unknown systematics: parameter estimation with theoretical uncertainty}
\label{sec:envelope}

Theoretical uncertainties and their applications in the context of sensitivity forecasts have been studied in previous work (e.g.~\citealt{Audren2013NeutrinoErrors,Baldauf2016,Sprenger2019}). In this work, we propose to extend the use of theoretical uncertainties in order to mitigate the negative impact of theory and code disagreements on parameter estimations due to unknown systematic effects (see \autoref{subsec:codedisagreementimpact}). To this end it is natural to start with a theoretical uncertainty readily available such as the one published in \citet{Sprenger2019} and study its effects on parameter estimation. We hence review this particular uncertainty envelope in \autoref{subsubsec:SprengerUncertainty}. In addition we introduce a new theoretical uncertainty in \autoref{subsubsec:KBUncertainty} that is tailored to the power spectrum prediction codes used in this paper. The comparison of the two theoretical uncertainties is presented later in the context of a mock parameter estimation in \autoref{subsec:tue}.

\subsubsection{Review of the simulation-based theoretical uncertainty}
\label{subsubsec:SprengerUncertainty}
We start out reviewing the theoretical uncertainty of~\citet{Sprenger2019}, which aims to identify the theoretical precision with which we will be able to analyze {\Euclid} data using cosmological simulations. This uncertainty envelope, in the following denoted $\sigma_{\mathrm{th, S}}$, is roughly given by the agreement between hydrodynamical simulations at the time of publishing of that work, which is expected to improve by the time {\Euclid} data is available and being analyzed. The envelope $\sigma_{\mathrm{th, S}}$ may be sufficient for any traditional forecasting with an aim to predict the sensitivity of future surveys as state-of-the-art tools on average disagree at a level that is mostly captured by $\sigma_{\mathrm{th, S}}$~\citep{Sprenger2019}.

The approach of \citet{Sprenger2019} takes three uncertainty values at three different wave numbers as inputs,
\begin{itemize}
    \item 0.33\% at $k=0.01 \hompc$\,,
    \item 1.0\% at $k=1 \hompc$\,,
    \item 10.0\% at $k=10 \hompc$
\end{itemize}
and interpolates between these points using the ansatz for the uncertainty envelope
\begin{equation}
    \label{eq:envelope}
    \alpha(k,z)=\begin{cases}a_1\exp{c_1\log_{10}\left[\frac{k}{k_{1\hompc}(z)}\right]},&\frac{k}{k_{1\hompc}(z)}<0.3\\a_2\exp{c_2\log_{10}\left[\frac{k}{k_{1\hompc}(z)}\right]},&\frac{k}{k_{1\hompc}(z)}>0.3\end{cases}\,.
\end{equation}
where $\alpha\propto\sigma_{\mathrm{th, S}}$ and $k$ depends on the line of sight direction $\mu$ and the cosmology, $k = k(\mu, \mathrm{cosmo})$ (for more details on this point see \citealt{Sprenger2019}).

In general, we define 
\begin{equation}
    \label{eq:zscaling}
    k_{k_0}(z) = k_0 \cdot (1+z)^\frac{2}{2+n_{\rm{s}}}\,.
\end{equation}
such that $k_{1\hompc}(z)$ is defined accordingly with $k_0 = 1\hompc$. In order to interpolate between the points mentioned above the coefficients are set to 

\begin{equation}
    a_1 = 0.014806,\,
    a_2 = 0.022047,\, 
    c_1 = 0.75056,\,
    c_2 = 1.5120\,. 
\end{equation}
See \autoref{fig:theoruncertainty} for an illustration of the uncertainty envelopes, including their redshift dependence. The blue lines represent $\sigma_{\rm{th, S}}$ at three different redshifts.

In \cite{Sprenger2019} (see their \autoref{sec:parameterestimation} for details) the theoretical $1\sigma$ uncertainty was given as
\begin{equation}
    \sigma_{\rm{th}}(k,\mu,z) = \left[\frac{V_r(z)}{2(2\pi)^2}k^2\Delta k \Delta \mu \frac{\Delta z}{\Delta \bar{z}}\right]^{1/2} \alpha(k,\mu,z) P_{\rm{gg}}(k,\mu,z)
\end{equation}
with redshift correlation length $\Delta z$, survey redshift bin width $\Delta \bar{z}$, survey volume per redshift bin $V_r(z)$, and wave number correlation length $\Delta k$. This uncertainty is then simply added in quadrature to the observational uncertainty in the $\chi^2$ computation.

However, we note that the publicly available likelihood (as of MontePython version 3.4) has a bug resulting in an inverted ratio of $(\frac{\Delta z}{\Delta \bar{z}})^{-1}$, which with the chosen redshift correlation length $\Delta z = 1$ and survey redshift bin width $\Delta \bar{z} = 0.1$ gives rise to a factor $(100)^{1/2}=10$ smaller uncertainty than intended. Since correlation lengths are inherently difficult to determine, and \cite{Sprenger2019} used very conservative correlation lengths, this can largely be interpreted as an adjustment to less conservative correlation lengths that would still be within the range of possible values: i.e. if we use $\Delta z = 0.1$ instead of $\Delta z = 1$, $\Delta \mu = 1$ instead of $\Delta \mu = 2$, and $\Delta k = 0.02 \hompc$ instead of $\Delta k = 0.05 \hompc$ then without the bug we would have a factor 2 smaller uncertainty than intended. Although unfortunate, given that the precision of future cosmological simulations that the uncertainty is intended to mimic is not known and~\cite{Sprenger2019} instead used the precision near the time of writing, a factor 2 smaller uncertainty (improved precision) is not unrealistic.

In contrast to the hydrodynamical simulations mentioned above, {\Halofit} and {\HMCode} exhibit an error on {\BAO} scales that is sufficient to bias the result of data analyses\footnote{Note this specific issue was mitigated in the latest version of {\HMCode} \cite{Mead2021}}. While we would expect the Euclid Collaboration analysis of future data to employ methods that accurately capture the {\BAO} scales, leaving the \cite{Sprenger2019} theoretical uncertainty robust for forecasts, we find in this paper (cf. \autoref{subsec:code_dependence}) that the bias remains for analyses that employ commonly used current fast methods such as e.g. {\Halofit} and {\HMCode}. Reducing this bias motivates the construction of a new uncertainty envelope that will be the topic of the next subsection.

\subsubsection{Theoretical uncertainty based on non-linear predictors}
\label{subsubsec:KBUncertainty}
\begin{figure}
	\includegraphics[width=\columnwidth]{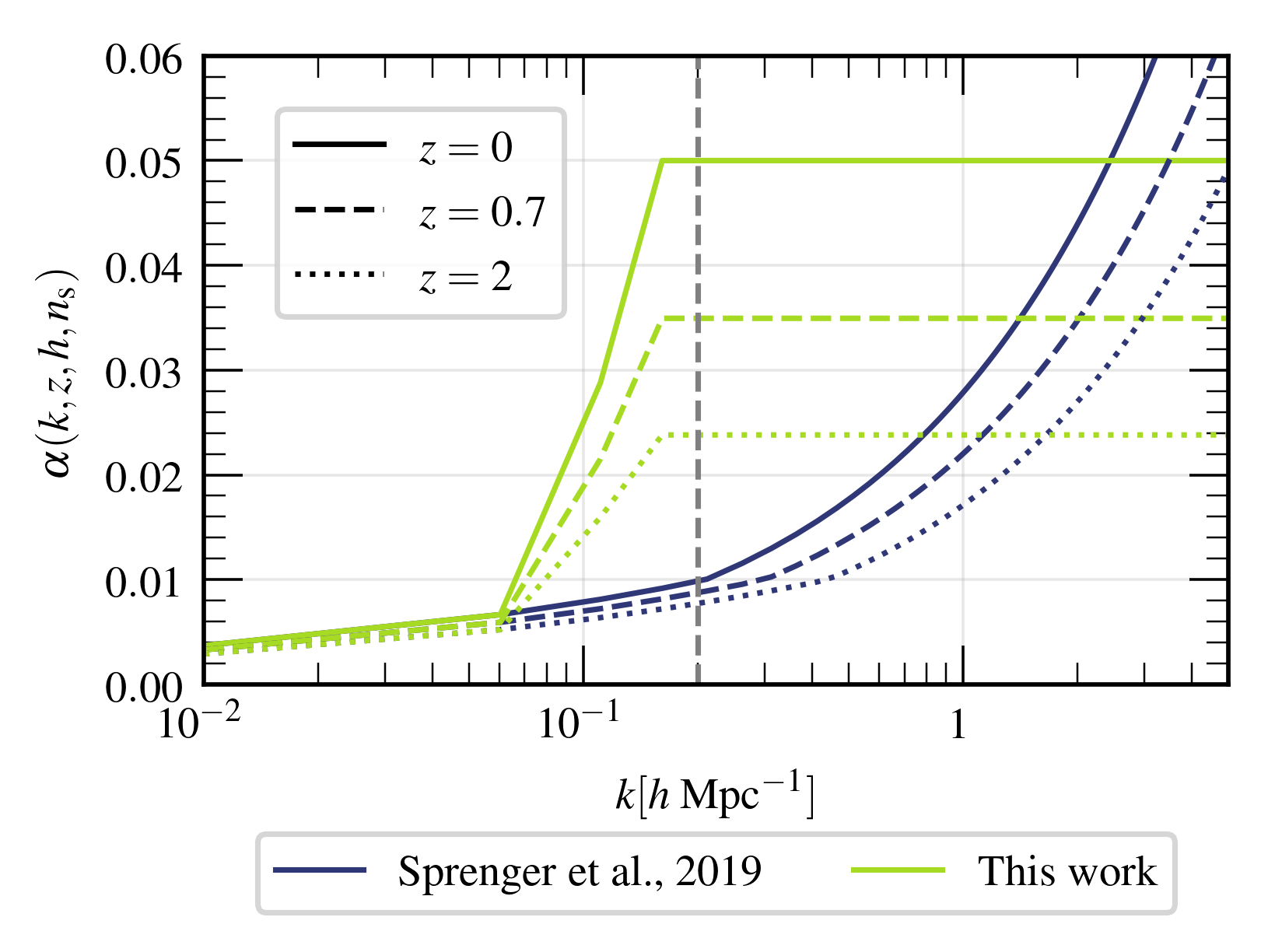}
	\caption{The two different envelope functions $\alpha$ for theoretical uncertainty as defined in \autoref{eq:envelope}. The solid, dashed and dotted lines indicate the uncertainty envelope at redshifts $z=0$, $z=0.7$ and $z=2$, respectively, where $[0.7,2]$ is the redshift range of the Euclid likelihood. Blue lines indicate the original theoretical uncertainty $\sigma_{\rm{th, S}}$ from \citet{Sprenger2019} and green lines correspond to $\sigma_{\rm{th, KB}}$, a rough estimate of the agreement between {\EEone} and {\HMCode}, as described in the text in \autoref{sec:envelope}. The vertical dashed grey line indicates $k=0.2\hompc$ which is a popular choice for the upper $k$ cut-off value. In most cases, we use the theoretical uncertainty $\sigma_{\rm{th, S}}$, but at the end we compare to the other uncertainty for a cross-comparison case.}
    \label{fig:theoruncertainty}
\end{figure}

{\EEone} has been compared to {\Halofit} and {\HMCode} in \autoref{fig:var} (for a comparison between {\EEone} and {\Halofit} also see \citealt{Knabenhans2019}). The agreement at linear scales is almost perfect. At intermediate scales the agreement is at the 1-2\% level for redshift $z=2$, while it is at the few percent level for $z=0$. For smaller scales the agreement is at the few percent level for all redshifts $0 \leq z \leq 2$. We therefore construct a new theoretical uncertainty envelope $\sigma_{\rm th, KB}$ based on the disagreement between these codes (see green lines in \autoref{fig:theoruncertainty}).

We find that this new uncertainty envelope reduces the bias on (mock) data analyses significantly (discussion in \autoref{subsec:tue}). However, while the disagreement between {\Halofit}/{\HMCode} and {\EEone} oscillates considerably in $k$ at intermediate and small scales for all $z$, for the sake of simplicity we chose to model the uncertainties with a constant close to the maximal disagreement found on scales $0.1\hompc\lesssim k\lesssim 1\hompc$. Hence, at these scales our uncertainty is somewhat overly conservative. The resulting uncertainty envelope follows the one by \citet{Sprenger2019} up to $k=0.05 \hompc$ and is fixed to 5\% above $k=0.15 \hompc$ at $z=0$. The intermediate part is modelled by \autoref{eq:envelope} using the coefficients $a_2 = 0.2998$ and $c_2 = 2.2517$, with the amplitude fixed to $\sigma_{\rm th, S}$ at $k=0.05 \hompc$.

Another question is up to which $k_{\rm max}$ mode these uncertainties can be trusted. Such a $k_{\rm max}$ represents a cut-off scale above which the theoretical uncertainty is de facto infinite. In other words, all information coming from $k$ modes larger than $k_{\rm max}$ is simply discarded. We use two different cut-off points: one ultra-conservative disregarding all scales above $k_{\rm max}=k_{0, \rm max}=0.2\hompc$ at $z=0$, and one informed by our choice of galaxy bias model (see~\autoref{sec:galaxypower}), which is only reliable up to about $k_{\rm max}=k_{0, \rm max}\approx0.4\hompc$ at $z=0$. Both scale with redshift as defined in \autoref{eq:zscaling}.

%% file: Chapters/03MCMCs.tex
\section{Analysis methodology}
\label{sec:analysis}
In \autoref{sec:introduction}, we listed the four key questions that guide our efforts in this paper:
\begin{enumerate}
  \item What are the performance differences of {\EEone}, {\Halofit} and {\HMCode}?
  \item What is the added value of considering (mildly) nonlinear scales in parameter forecasts?
  \item What is the impact of the choice of the (non-linear) predictor model on the parameter estimation result?
  \item How are the forecasting results affected by different choices of theoretical uncertainty models?
\end{enumerate}
In order to answer these questions we run and analyze a set of {\MCMCs}. In this section we discuss the general aspects of the {\MCMC} analysis methodology as well as the experimental setup used to generate those {\MCMCs}. The results and interpretations will be reported later in \autoref{sec:forecasts} and \ref{sec:parameterestimation}.

\subsection{Modelling the galaxy-galaxy power spectra}
\label{sec:galaxypower}
{\EEone} predicts the non-linear correction to the dark matter power spectrum only, which is not an observable quantity. A related quantity that we can measure from observations is the galaxy-galaxy power spectrum $P_{\rm gg}(k,\mu,z)$, where $\mu$ is the cosine between the wave mode $\vk$ and the line of sight direction $\ve_r$. Following the expansion in \cite{Sprenger2019} and using their notation we have
\begin{equation}
\begin{split}
    P_{\rm gg}(k,\mu,z) =&\quad f_{\rm AP}(z)\times f_{\rm res}(k,\mu,z)\times f_{\rm RSD}(\hat{k}, \hat{\mu}, z)\times b^2(z)\\
    &\times P_{\delta\delta}(k,z)\,,
\end{split}
\label{eq:PggExpansion}
\end{equation}
where $f_{\rm AP}$ denotes the correction due to the Alcock-Paczy\'nski effect, $f_{\rm res}$ is a correction factor necessary because the resolution is limited, $f_{\rm RSD}$ accounts for the {\RSDs} and $b(z)$ is the galaxy bias taking into account that galaxies are only biased tracers of the underlying matter distribution. Notice that the parameters $k$ and $\mu$ are actually functions of the underlying cosmology. Parameters with a hat denote that they are based on the true/real but unknown cosmology, while those without a hat are based on an assumed underlying cosmology. For a more elaborate explanation of this see \citet{Sprenger2019}.

For the galaxy bias we choose a linear, redshift-dependent bias $b(z) = \beta_0(1+z)^{0.5 \beta_1}$ as in~\cite{Sprenger2019}, where $\beta_0$ and $\beta_1$ are included as nuisance parameters in the MCMC analysis. Since galaxy bias is known to be better described by a scale-dependent galaxy bias (e.g. \citealt{Desjacques2018, Giusarma2018}), we choose to cut the analysis at an ultra-conservative $k_{\rm{max}}=0.2\hompc$ and a conservative $k_{\rm{max}}=0.4\hompc$, assuming that any residual inaccuracy of the bias model up to these scales would be picked up by the theoretical uncertainty.

The peculiar motion of galaxies give rise to the so-called Fingers of God \citep{1978IAUS...79...31T,Jackson:2008yv} and are treated separately as an exponential suppression of power following~\cite{Bull:2014rha}, where the suppression is governed by a nuisance parameter $\sigma_{\rm NL}$ that we marginalize over. The remaining {\RSD} effect is described by the Kaiser formula \citep{Kaiser1987ClusteringSpace}, which relates the galaxy power spectrum $P_{\rm gg}$ to the dark matter power spectrum $P_{\delta\delta}$. {\EEone} directly gives us an {\NLC} factor $B(k,z) = P_{\delta\delta}^{\rm non-linear}(k,z) / P_{\delta\delta}^{\rm linear}(k,z)$, which, setting aside the other correction factors for the moment, allows us to write the {\RSD} and bias correction as (see~\autoref{app:biasmodel})
\begin{equation}
\label{eq: nonlinear_kaiser1}
   P_{\rm gg}(k,\mu,z) = b^2(z)\left[B(k,z)+2\mu^2f+\mu^4f^2\right]P_{\delta\delta}^{\rm linear}(k,z)\,.
\end{equation}
For {\Halofit} and {\HMCode}, which return a non-linear power spectrum, we find identical results expressing this as (see~\autoref{app:biasmodel})
\begin{equation}
\label{eq: nonlinear_kaiser2}
     P_{\rm gg}(k,\mu,z) = b^2(z)\left[1+\mu^2f\right]^2P_{\delta\delta}^{\rm non-linear}(k,z)
\end{equation}
The other correction factors are multiplicative and can trivially be included in these expressions. For deeper discussion of all of these correction factors we refer the reader to \cite{Sprenger2019} and references therein.

\subsection{Experimental setup}
We run the \gls{MCMC}s with MontePython version 3.1 \citep{Audren2013ConservativePYTHON,Brinckmann2019} and CLASS version 2.8 \citep{Blas2011,Lesgourgues2011a}. We use the publicly available Euclid galaxy clustering likelihood\footnote{Note that in cosmological inference the term "likelihood" is typically a package containing all modelling of observables plus the likelihood computation.} from~\citet{Sprenger2019}, modified according to \autoref{sec:galaxypower} and \autoref{sec:envelope} to allow for including non-linear corrections by the {\EEone} and a modified theoretical uncertainty scheme.

The likelihood uses 13 redshift bins in the redshift range $0.7<z<2$, a sky fraction $f_{\rm{sky}} = 0.3636$, nine $\mu$ bins, and 100 logarithmically spaced bins in $k$-space from $k_{\rm{min}}=0.02\,\text{Mpc}^{-1}$ to $k_{\rm{max}}=0.2$~or~$0.4\hompc$ (see the end of~\autoref{sec:envelope} for a discussion of $k_{\rm{max}}$). For more details we refer the reader to~\citet{Sprenger2019}.

We combine the Euclid likelihood with the fake Planck likelihood and noise spectra from~\citet{Brinckmann2019TheMeasurement}, which is based on the old "fake planck bluebook" likelihood with specifications updated to the full mission and omitting the 217 GHz channel, in order to avoid foreground complications. The end result is a likelihood in good agreement with the sensitivity of the Planck 2018 results~\citep{Aghanim:2018eyx}, but with a slightly better sensitivity for the optical depth to reionization, $\sigma(\tau_{\rm{reio}})$. The fake Planck likelihood considers a maximum multipole of $\ell_{\rm{max}}=3000$ for temperature, polarisation and lensing, with the noise exploding long before the maximum multipole (see Figure~1 of~\citealt{Brinckmann2019TheMeasurement}).

\begin{table}[H]
\centering%
\caption{Cosmological (top panel) and nuisance parameters of the fiducial cosmology used for all \gls{MCMC} forecasting runs. The middle panel lists the varying RSD and galaxy bias nuisance parameters of the {\Euclid} likelihood (see \autoref{sec:galaxypower}), while the bottom panel contains baryonic feedback nuisance parameters of {\HMCode}, which remain fixed. The fiducial cosmology coincides with the Euclid Reference Cosmology (as defined in the \citet{Knabenhans2019}. Notice that the minimal and maximal values are broadly equal to the parameter box of {\EEone} (keep in mind that $\Omega_{\rm cdm}h^2$ is not a direct input parameter to {\EEone}).}
\begin{tabular}{llll}
Parameter & fiducial value & min & max\\
\hline
$\Omega_{\rm b}h^2$ & $0.022$ & $0.0215$ & $0.0235$\\
$\Omega_{\rm cdm}h^2$ & $0.121$ & $0.1091$ & $0.1311$\\
$n_{\rm{s}}$ & $0.96$ & $0.9283$ & $1.0027$\\
$h$ & $0.67$ & $0.6155$ & $0.7307$\\
$w_0$ & $0.9619$ & $-1.25$ & $-0.75$\\
$\sigma_8$ & $0.83$ & $0.7591$ & $0.8707$ \\
$z_{\rm reio}$ & $8.24$ & $0.0$ & none\\
\hline
$\sigma_{\rm NL}$ & $7.0$ & 4.0 & 10.0\\
$\beta_0^{\rm Euclid}$ & $1.0$ & none & none\\
$\beta_1^{\rm Euclid}$ & $1.0$ & none & none\\
\hline
$\eta_0$ & 0.603 & \multicolumn{2}{c}{fixed} \\
$c_{\rm{min}}$ & 3.13 & \multicolumn{2}{c}{fixed} \\
\hline
\hline
\label{tab:FidCosmo}
\end{tabular}
\end{table}

\subsection{MCMC analysis}
We use the {\MCMC} technique to compare the agreement between a fiducial and a predicted galaxy-galaxy power spectrum, $P_{\rm{gg}}$, where the latter takes a cosmology as an input argument. While $P_{\rm{gg}}$ is the observable galaxy clustering quantity used in this work, the three codes compared to one another all predict the non-observable matter power spectrum $P_{\delta\delta}$. Technical details about how $P_{\rm{gg}}$ is related to the $P_{\delta\delta}$ and how the {\NLC} predicted by {\EEone} is worked into this relation are given in \autoref{eq:PggExpansion} and \autoref{eq: nonlinear_kaiser1}. 

In this paper, we perform sensitivity forecasts and mock parameter estimations instead of real-world data analysis. Hence, we first have to construct a fiducial model. As will be explained in more detail below, our \gls{MCMC} runs can be split into two classes, which we refer to as ``auto-comparison'' (ac) and ``cross-comparison'' (cc) tests (see \autoref{subsec:ac_vs_cc} for an explanation why we do this). In the case of auto-comparisons, the fiducial model is set up with the same cosmological model, and cut-off scale as the model used to perform the theoretical predictions during the \gls{MCMC}. For cross-comparisons, on the other hand, we compute a fiducial spectrum with {\EEone} in all cases and use either {\Halofit} or {\HMCode} for the theoretical predictions. Notice that the choice of the theoretical uncertainty model does not affect the computation of the fiducial result.

We let the \gls{MCMC}s run on the zBox4 cluster based at the University of Zurich\footnote{\url{https://www.ics.uzh.ch/~stadel/doku.php?id=zbox:zbox4}}. We run twelve chains simultaneously for each \gls{MCMC}. For any information about how this works in detail, we refer the reader to \citet{Audren2013ConservativePYTHON, Brinckmann2019}.

To determine convergence we use the standard criterion by \citet{Gelman1992InferenceSequences}, $R-1<0.01$, where $R$ is the Gelman-Rubin coefficient\footnote{Note that the accuracy of the Gelman-Rubin convergence criterium depends on the number of chains as well as the length of the chains, but with 12 chains per run $R-1<0.01$ leads to exceedingly good convergence.}. We let the \gls{MCMC}s run until this convergence criterion is satisfied in all parameters.

%% file: Chapters/04Forecasts.tex
\section{Sensitivity forecasts}
\label{sec:forecasts}
Traditionally, different power spectrum predictors are compared at the power spectrum level as is done e.g. in \citet{Lawrence2010, Bird2012, Mead2016, Lawrence2017, Knabenhans2019, Angulo2020, Knabenhans2021, Mead2021}. However, it is not often studied directly how the differences between the predictors impact parameter forecasts in the end (see, however, \cite{Martinelli2021} for a complementary analysis for cosmic shear). We close part of this gap in this section by analyzing the code dependence of forecasting results on the three codes studied in this paper, leaving investigations of this aspect with other codes to future work.

Code dependencies in forecasting results could manifest themselves in different ways: uncertainty contours could be smaller or show different degeneracy directions. Ultimately, we would want to understand the reasons behind any differences, which could be due to mistakes in the modelling, e.g. underestimating uncertainties, not accurately modelling uncertainties (e.g. due to baryonic effects), or incorrectly modelling non-linear effects, or it could be due to new physical signals, e.g. more accurate modelling of non-linear effects introducing new signals that were neglected with other methods.

In this section, we perform an auto-comparison \gls{MCMC} for each code in which we try to fit a fiducial power spectrum with the same predictor code as the one used to compute the fiducial\footnote{This is the standard approach to forecasting and has been shown to provide reliable uncertainties compared to a randomly generated fiducial, as long as the difference cosmology is not extreme~\citep{Perotto2006ProbingSimulations}.}. In such a comparison, no significant difference in the best-fit cosmologies between the different codes is expected and differences in the posterior probabilities are limited to different sensitivity contours.

\subsection{Sensitivity to the cosmological parameters}
\label{subsec:results_auto}
We first compare the sensitivity to the cosmological parameters for Euclid galaxy clustering plus (fake) Planck {\CMB} for each of the three different predictors looked at in this paper. To this end, we analyze four different cases: {\LCDM} and {\wCDM} as well as two different cut-off scales, $k_{\rm{max}} = 0.2\hompc$ and $k_{\rm{max}} = 0.4\hompc$.

\begin{figure}
	\includegraphics[width=\columnwidth]{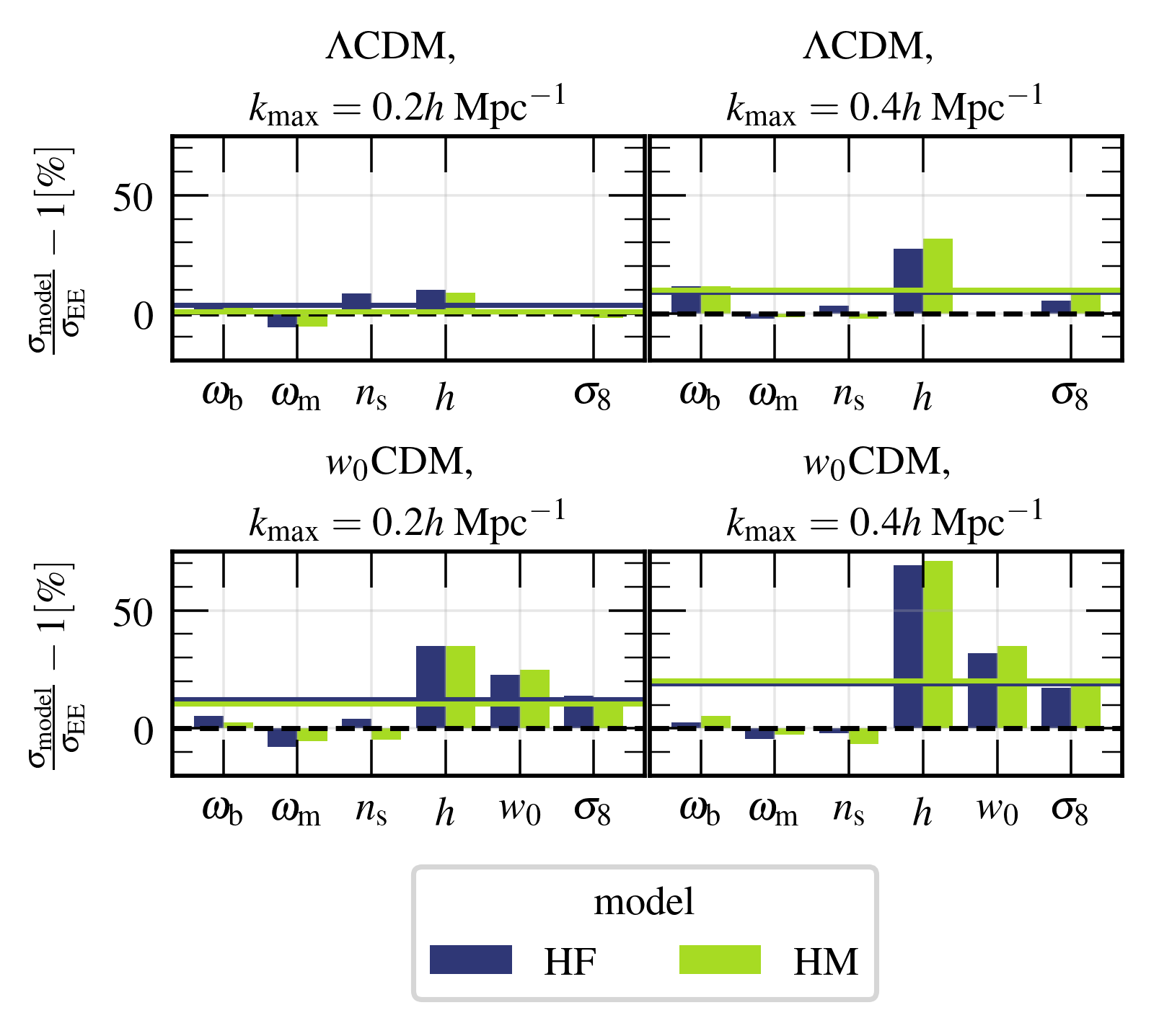}
	\caption{Relative differences (in \%) of 1D marginalized posterior standard deviations ($\sigma$) for each cosmological parameter between the two models {\Halofit} (HF) and {\HMCode} (HM), respectively, with respect to {\EEone} (EE) for all four tested cases (given in the panel titles). The solid horizontal lines represent the average over all parameters. Notice that a positive ratio implies that $\sigma_{\rm EE} < \sigma_{\rm{model}}$. In all four cases, on average, the 1D marginalized posterior standard deviations computed with {\EEone} are smaller than those computed with either of the two alternative models. The more complex the cosmological model and also the higher the cut-off scale $k_{\rm max}$, the bigger the effect, i.e. the smaller the uncertainty obtained from a {\EEone}-based fit compared to a {\Halofit}- or {\HMCode}-based fit. The averaged relative differences vary between $0.9\%$ ({\HMCode} vs {\EEone}, {\LCDM}, $k_{\mathrm{max}}=0.2\hompc$) and $20.4\%$ ({\HMCode} vs {\EEone}, {\wCDM}, $k_{\mathrm{max}}=0.4\hompc$) depending on the case.}
    \label{fig:ac_sigma_analysis}
\end{figure}

The 1D marginalized posterior standard deviations for each parameter in all four cases and for all three predictors are summarized in \autoref{tab:autoconvres}. The relative differences of these standard deviations are graphically represented in \autoref{fig:ac_sigma_analysis} for better understanding. 

\begin{figure}
	\includegraphics[width=\columnwidth]{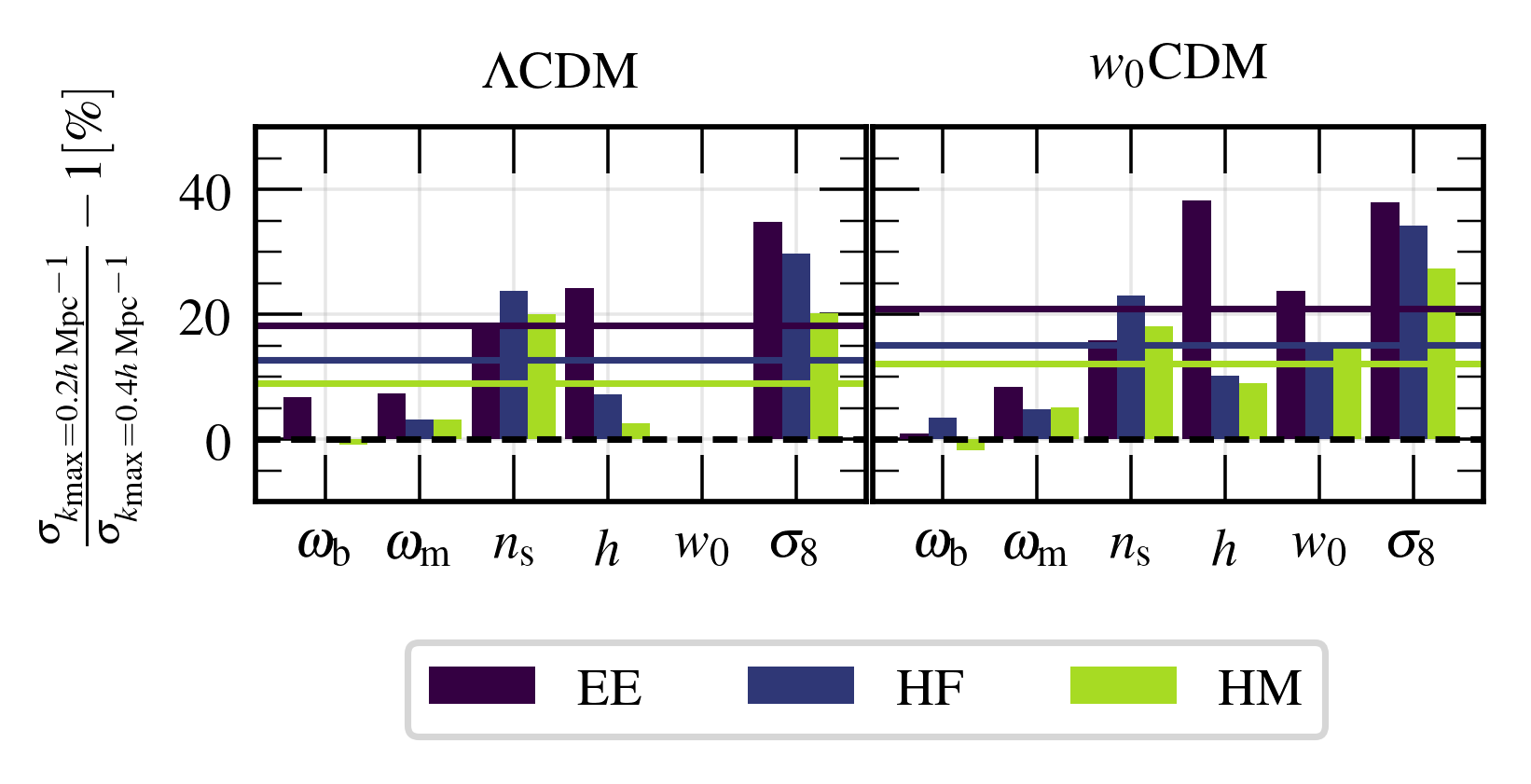}
	\caption{Relative differences (in \%) of 1D marginalized posterior standard deviations ($\sigma$) per cosmological parameter between the two different cut-off scales $k_{\rm max}=0.2\hompc$ and $k_{\rm max}=0.4\hompc$ for each of the three predictors. The averaged relative differences range between $9.2\%$ and $18.2\%$ for {\LCDM} and from $12.1\%$ up to $20.9\%$ in the {\wCDM} case. From this plot it becomes most evident that including smaller scales leads to smaller uncertainties.}
    \label{fig:kmax_effect}
\end{figure}

Clearly, this relative difference varies dramatically depending on the considered parameter: for instance, the sensitivity of {\EEone} to {\omm} in the case of a {\LCDM} cosmology cut off at $k_{\mathrm{max}}=0.2\hompc$ is almost the same compared to the corresponding sensitivity of {\HMCode} (the relative difference is around $1\%$). However, $\sigma_\mathrm{EE}$ for $h$ in the case of a {\wCDM} cosmology cut off at $k_{\mathrm{max}}=0.4\hompc$ is reduced by $\sim70\%$ compared to the corresponding $\sigma_\mathrm{HM}$ value. In other words: {\EEone} is roughly 70\% more sensitive to $h$ than {\HMCode}. Evidently, the standard deviations computed with {\EEone} are always smaller than those computed with either {\HMCode} or {\Halofit}. The more complex the cosmological model and also the higher the cut-off scale $k_{\rm max}$, the bigger the effect, i.e. the smaller the uncertainty obtained from a {\EEone}-based fit compared to a {\Halofit}- or {\HMCode}-based fit. 

In the same \autoref{fig:ac_sigma_analysis}, we also show the average values of those relative differences over all parameters (for each studied case). These averaged values range from only $0.9\%$ ({\HMCode} vs {\EEone}, {\LCDM}, $k_{\mathrm{max}}=0.2\hompc$) to $20.4\%$ ({\HMCode} vs {\EEone}, {\wCDM}, $k_{\mathrm{max}}=0.4\hompc$) depending on the case. This quantity serves as a proxy for the volume (to the $1/d$ power, where $d$ is the dimension of the parameter space under investigation) of the $1\sigma$ credible region of the posterior distribution and thus for the sensitivity to the cosmological parameters. As is stated above and clearly visible in \autoref{fig:ac_sigma_analysis}, this average value is always positive, meaning that on average the 1D marginalized posterior standard deviations computed from a {\EEone}-based posterior is smaller than the corresponding value of a posterior based on either {\Halofit} or {\HMCode}. It also becomes clear, that the effect is bigger the more complex the cosmological model under consideration and the higher the cut-off scale $k_{\rm max}$.

{\EEone} models the matter power spectrum much more accurately than either of the halo model-based models, particularly for $0.1 \hompc\lesssim k \lesssim 5\hompc$, as is shown in \citet{Knabenhans2019}. However, it is not guaranteed that a more accurate model produces smaller posterior contours (only less biased ones). The fact that we still observe smaller contours here may be explained by the fact that {\EEone} models the {\BAOs} more accurately than the other two predictors, i.e. scales that carry lots of valuable information about the cosmology. It should, however, not be forgotten that {\EEone} does not take baryonic effects into account, while {\HMCode} can. But we note that we did not take them into account even for the {\HMCode} runs, instead fixing the nuisance parameters related to baryonic physics to the default values (see \autoref{tab:FidCosmo}).

\paragraph*{Answer to question \#1:} Although {\EEone} models the $P(k)$ more accurately than {\Halofit} or {\HMCode}, it also leads, on average, to better sensitivity for the cosmological parameters than the latter two codes, with the largest improvements seen for $h$ and $w_0$ as {\EEone} is better at modelling the effect of these parameters on the shape of the matter power spectrum.

\subsection{The effect of including mildly non-linear scales}
\label{subsec:kmax_uncertainties}

In order to establish the importance of mildly non-linear scales in cosmological parameter estimations we perform auto-comparison tests for two different cut-off scales: $k_{\rm{max}} = 0.2\hompc$ and $k_{\rm{max}} = 0.4\hompc$. Although all predictors used in this study could actually reach more strongly non-linear scales, we do not go beyond $0.4\hompc$ because we use only a very simple, linear galaxy bias model which is not valid at small scales. 

We quantify the effect of including mildly non-linear scales on the posterior sampling by comparing the sizes of the resulting contours. Of course, a more in-depth analysis employing a more sophisticated galaxy bias model needs to be conducted in order to draw conclusion on the effect of including strongly non-linear scales. As this paper is meant to convey a proof of concept, we leave these analyses to future work.

\begin{figure*}
	\includegraphics[width=0.495\textwidth]{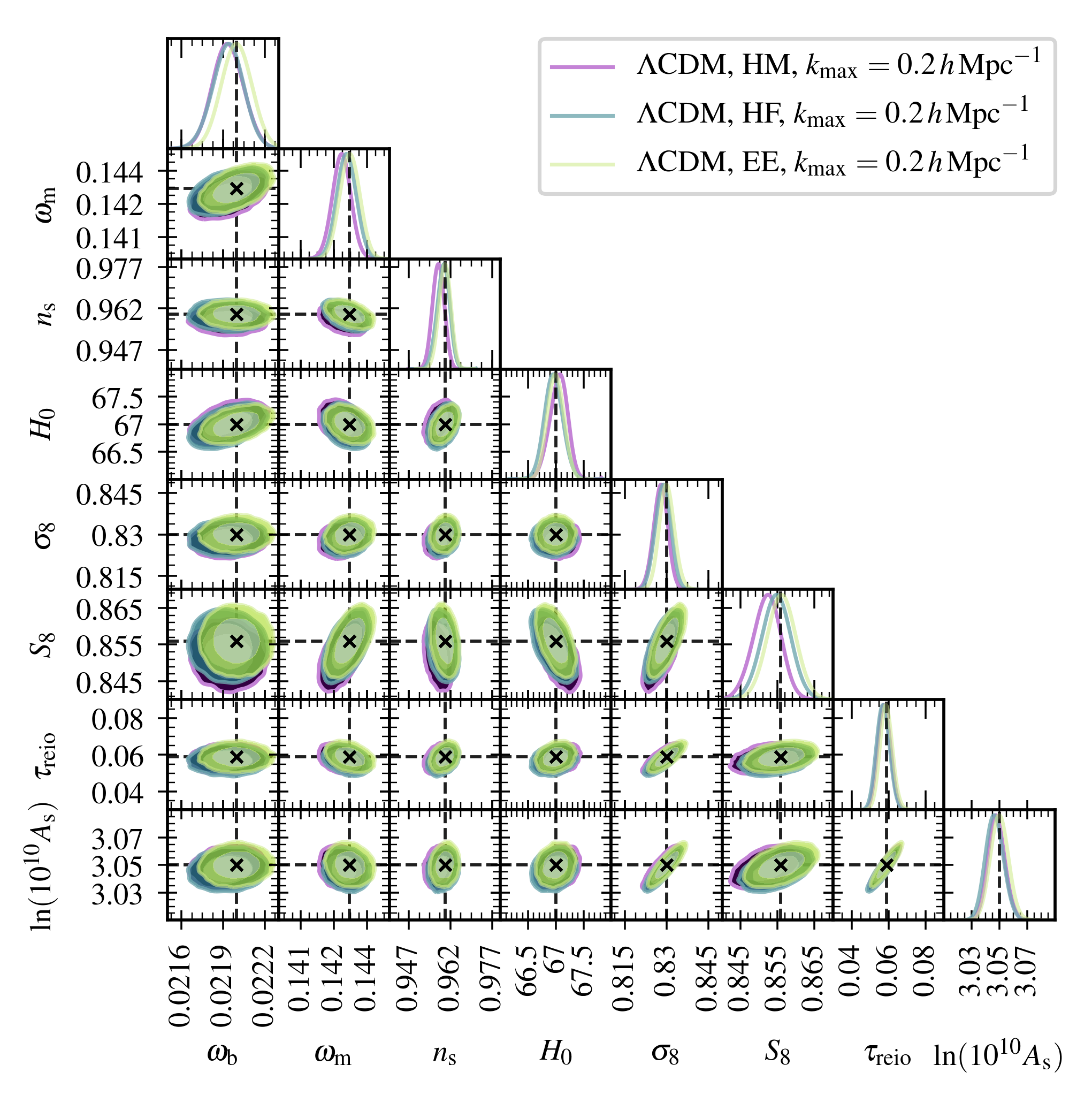}
	\includegraphics[width=0.495\textwidth]{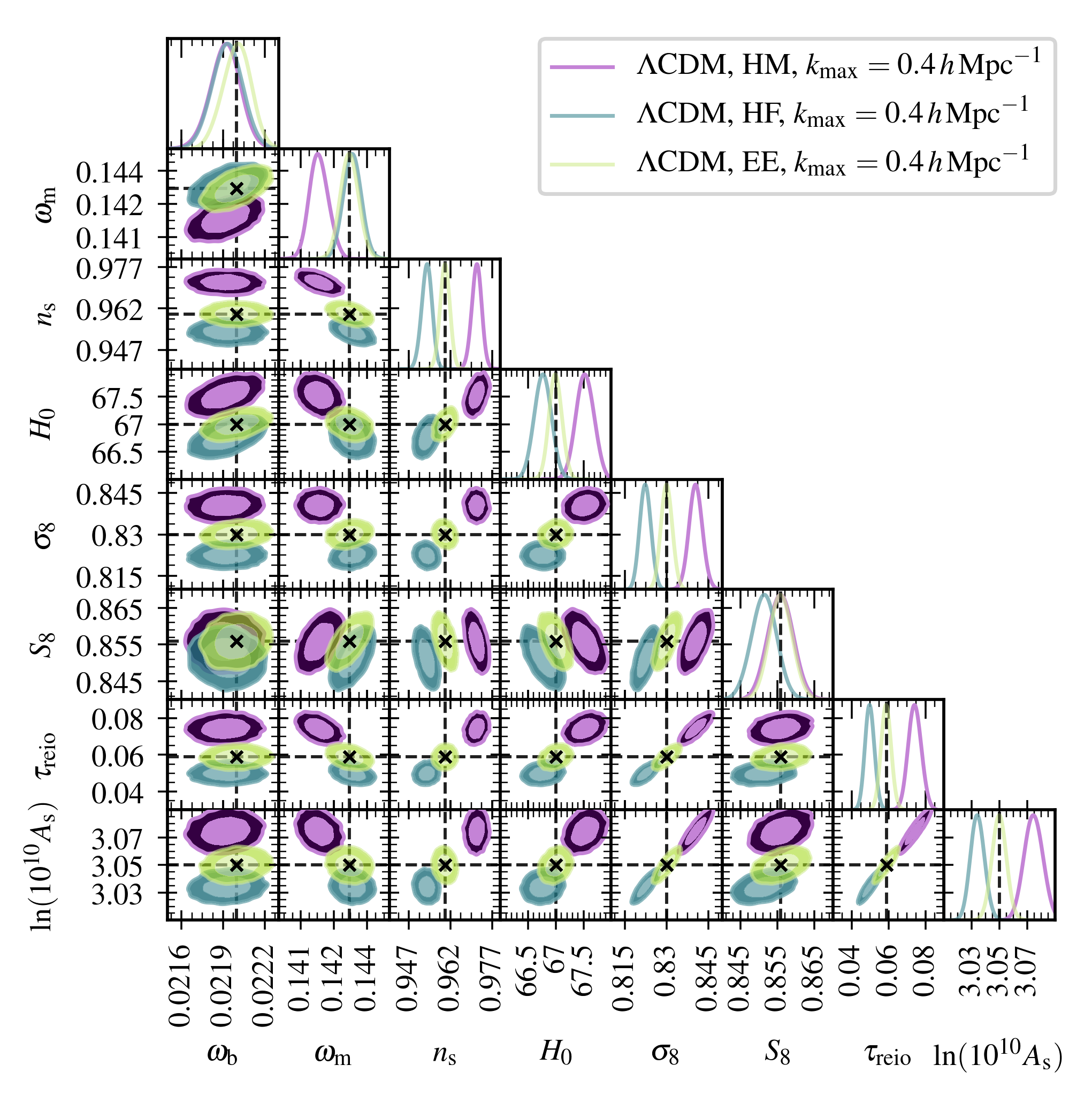}
	\caption{$\Lambda$CDM cross comparison of {\HMCode} and {\Halofit} vs {\EEone} considering $k\leq 0.2 \hompc$ (left triangle plot) and $k\leq 0.4 \hompc$ (right triangle plot), respectively. We find that all three best-fit cosmologies agree relatively well with each other in the left panel as mostly linear scales are considered. In the right panel, however, clear differences in the posterior distributions emerge due to disagreements in the power spectrum predictions by the three different codes.} 
	\label{fig:LCDM}
\end{figure*}

To this end, we interpret again the results presented in \autoref{tab:autoconvres}. This time, however, we shall not compare the different prediction models but perform all comparisons for each of the predictors separately. In \autoref{fig:kmax_effect} we plot the relative differences (again in \%) of the 1D marginalized posterior standard deviations between the two different cut-off scales for each of the power spectrum prediction models. Positive values imply that $\sigma_{k_{\rm max} = 0.4\hompc} < \sigma_{k_{\rm max} = 0.2\hompc}$. This is the case for almost all parameters for all predictors: the effect of increasing the cut-off scale is largest for {\EEone} where the posterior standard deviation averaged over all parameters is reduced by $18.2\%$ in the {\LCDM} scenario and by $20.9\%$ in the {\wCDM} case. The minimal effect found is $9.2\%$ ({\LCDM}) and $12.1\%$ ({\wCDM}), respectively. As above, we have represented the average effect (over all parameters) as solid horizontal lines. 

\paragraph*{Answer to question \#2:} The uncertainties can be significantly reduced by increasing $k_{\rm max}$ from $0.2\hompc$ to $0.4\hompc$. This is not only true for {\EEone} (though for it the effect is largest) but for all tested prediction models. The magnitude of the effect depends strongly on the model and on the predictor under consideration. This observation does not come as a surprise: it is expected that there is valuable information stored in the galaxy-galaxy power spectrum on (mildly) non-linear scales that can be leveraged in order to estimate the cosmological parameters. The effect is expected to grow as the cut-off scale increases. We reiterate, however, that increasing $k_{\rm max}$ beyond $\sim 0.4\hompc$ requires a more sophisticated galaxy bias model than is considered in this paper. So we leave this analysis to future work.

%% file: Chapters/05Estimation.tex
\section{Mock parameter estimation}
\label{sec:parameterestimation}
\label{subsec:ac_vs_cc}

Usually, for forecasting we use the same (non-linear) prediction model for cosmological parameter estimations as is used to compute the fiducial cosmology, i.e. auto-comparison tests are the common choice (like in the previous section). This is fine as long as we are interested in the sensitivity of the model response to the choice of parameters~\citep{Perotto2006ProbingSimulations}. We can, however, mimic the situation of data analysis by e.g. taking {\EEone} to be the model that produces the ``observed'' data which we try to fit with different predictors, which we refer to as the ``cross-comparison'' case. I.e., in this section, we always produce the ``observed'' mock data with {\EEone}. Running such cross-comparison tests using different predictors for the fitting are likely to lead to different best-fit cosmologies (as we will also confirm later), as different codes encode different physics and differ in the way they are implemented. Such tests can hence be used to emphasize that any cosmological parameter estimation result significantly depends on the code used, unless proper care is taken to account for uncertainties on non-linear scales.

\subsection{Code dependence of best-fit cosmology}
\label{subsec:code_dependence}

\begin{figure*}
    \includegraphics[width=0.495\textwidth]{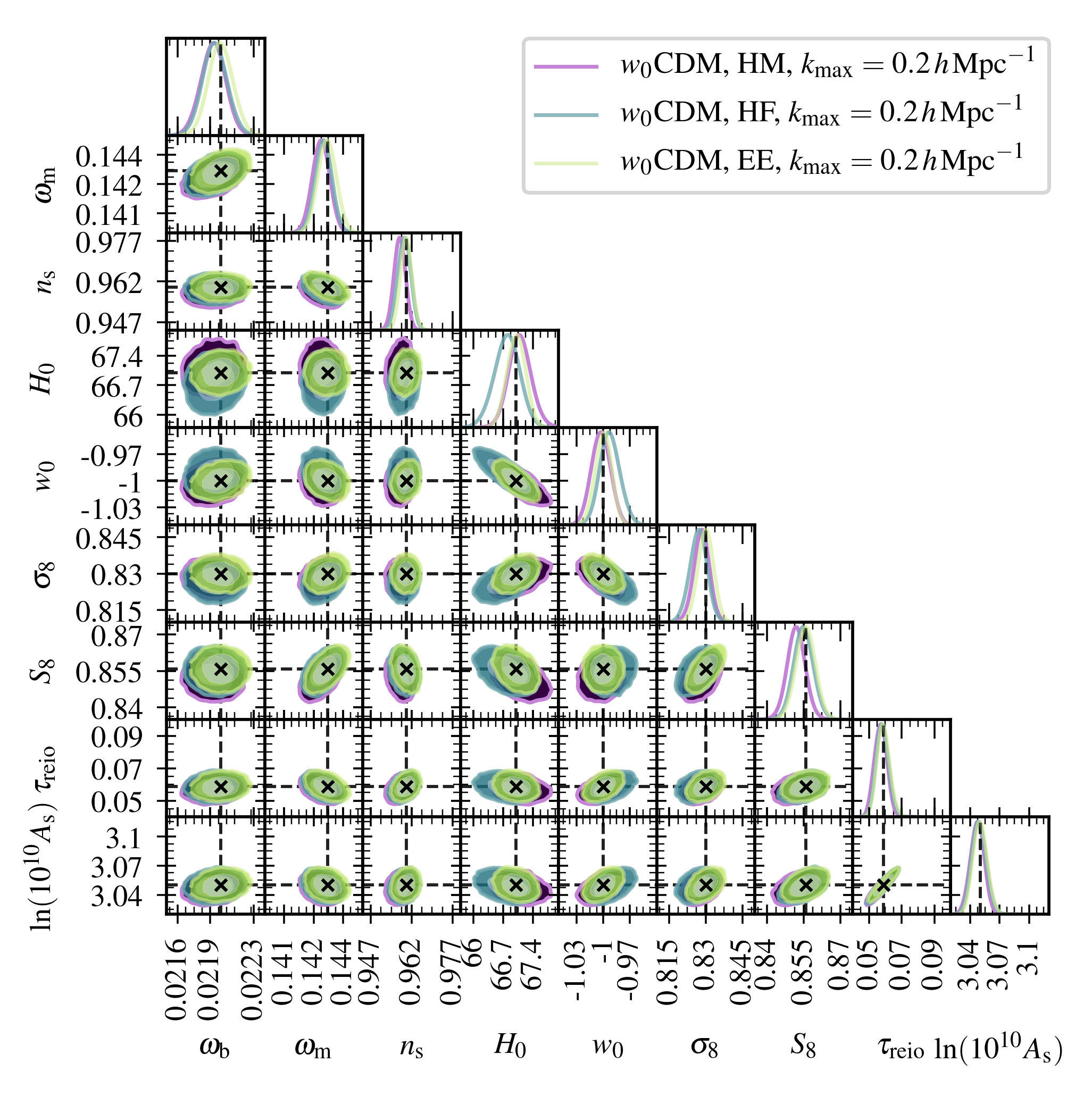}
    \includegraphics[width=0.495\textwidth]{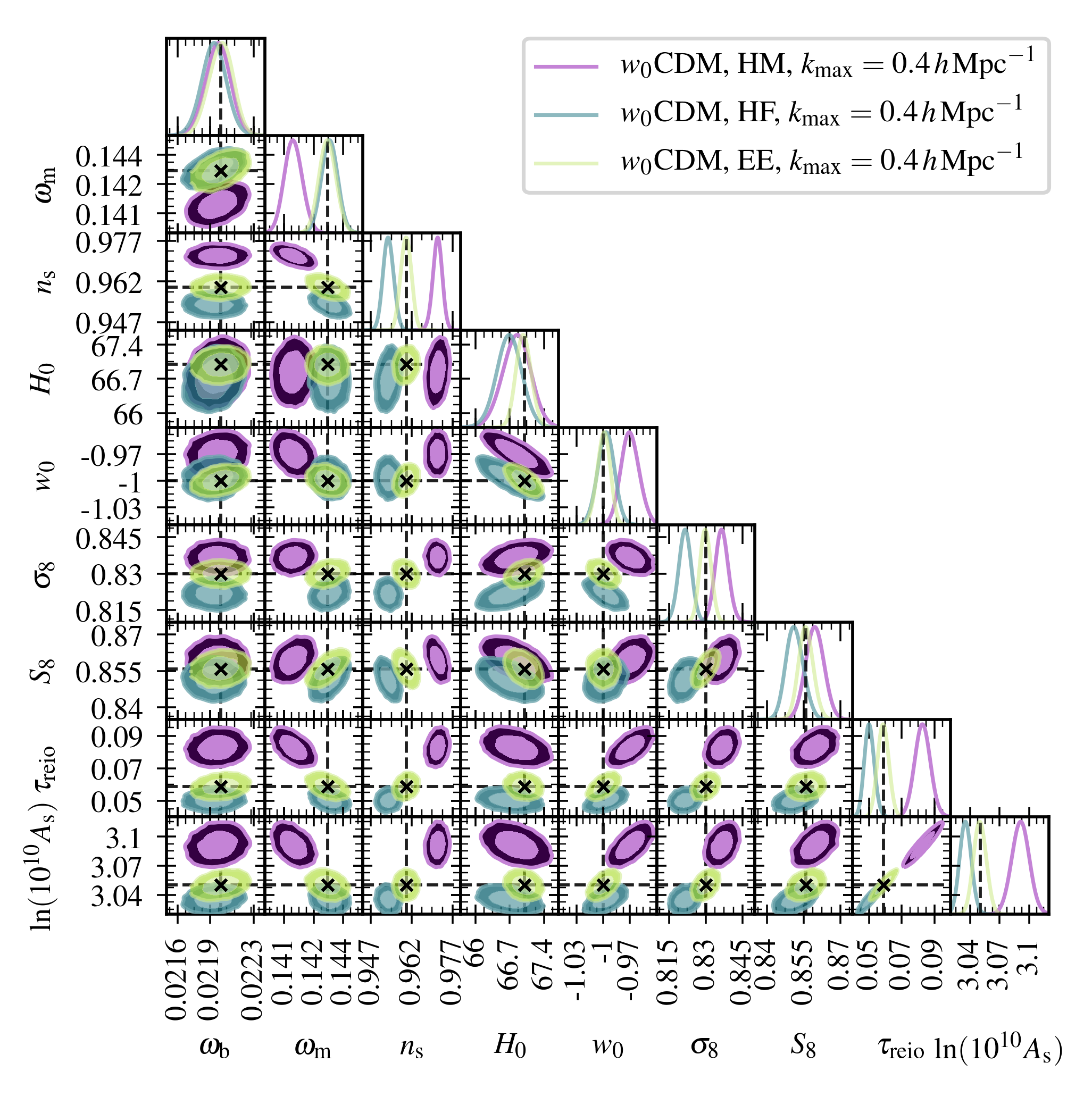}
	\caption{$w_0$CDM cross comparison of {\HMCode} and {\Halofit} vs {\EEone} considering $k\leq 0.2 \hompc$ (left triangle plot) and $k\leq 0.4 \hompc$ (right triangle plot), respectively. Just as in \autoref{fig:LCDM} we see good agreement between the posteriors in the case with the lower cut-off scale while increasing that scale leads to emergent differences. }
	\label{fig:wCDM_cc_0p4}
\end{figure*}

The third of our key questions (see \autoref{sec:analysis}) is concerned with the choice of the power spectrum prediction code and its effect on the best-fit cosmology. In this subsection, we analyze and interpret the results from cross-comparison tests, where we compare power spectrum fits computed with {\Halofit} and {\HMCode} with a fiducial power spectrum computed with {\EEone}. The resulting posteriors for the $\Lambda$CDM case are shown in  \autoref{fig:LCDM} (left panel: $k_{\rm max}=0.2\hompc$; right panel: $k_{\rm max}=0.4\hompc$) while the ones for the $w_0$CDM case are shown in \autoref{fig:wCDM_cc_0p4} (left panel: $k_{\rm max}=0.2\hompc$; right panel: $k_{\rm max}=0.4\hompc$). The numerical results for mean posterior cosmologies and the associated 68\% credible contours can be found in \autoref{tab:crossconvres}.

Looking at the figures showing the cases for $k_{\rm max}=0.2\hompc$, we can easily see that for both cosmological models, all codes largely agree on the best-fit cosmology if only $k\leq k_{\rm max}=0.2 \hompc$ are considered. More concretely, for these cases the fiducial cosmology always lies within the $1\sigma$ credible region. The uncertainties are broadly comparable, as only for $h$ (or equivalently $H_0$) is the sensitivity of {\EEone} considerably better than that of the other two codes. This agreement is not surprising at all, remembering that the three considered power spectrum prediction codes agree almost perfectly on linear scales \citep{Knabenhans2019, Knabenhans2021}.

\begin{figure}
	\includegraphics[width=\columnwidth]{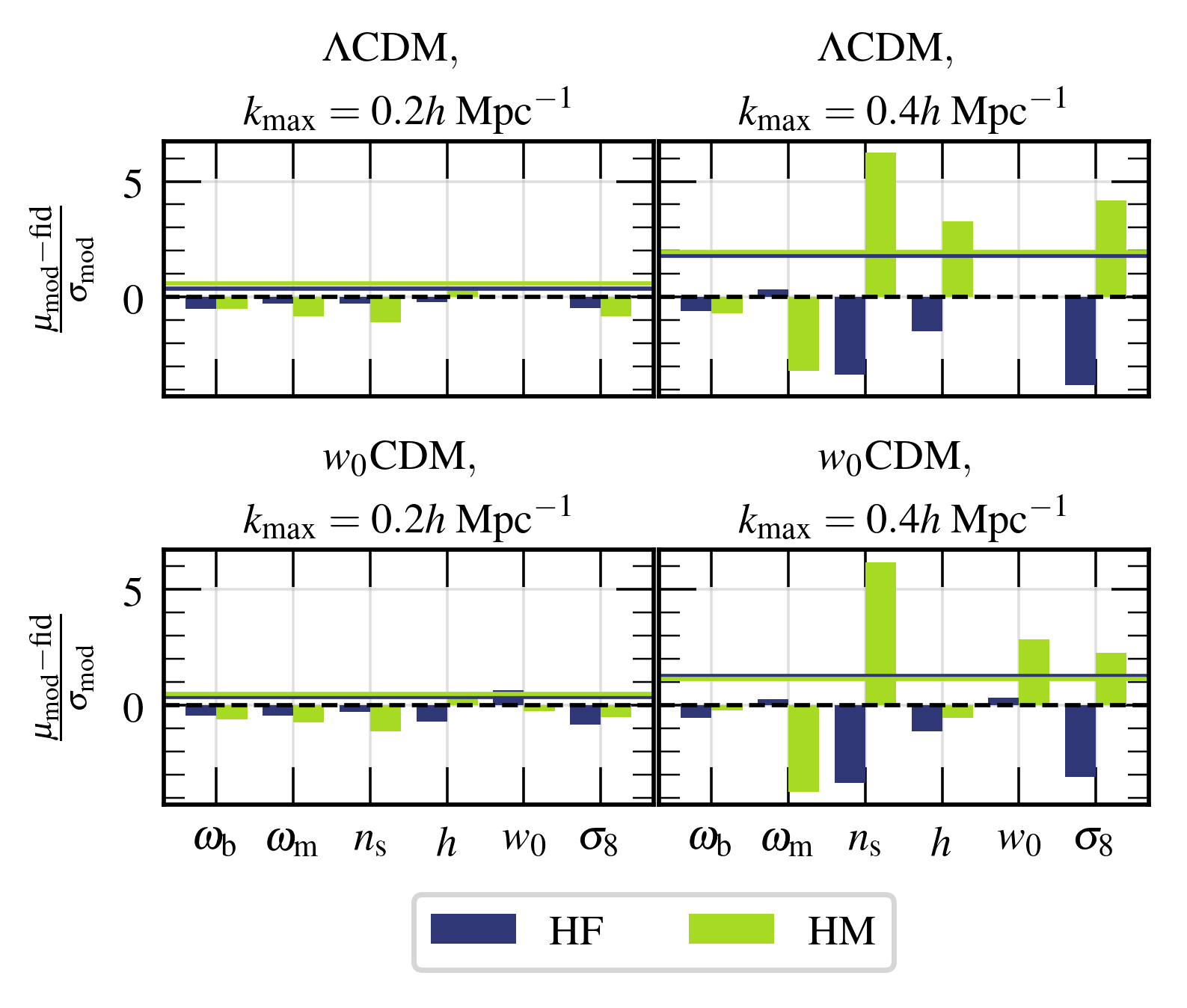}
	\caption{Differences between mean ($\mu$) and fiducial cosmology per cosmological parameter for {\Halofit} and {\HMCode} in the cross-comparison setup and for both cut-off scales. The differences are given in units of the respective 1D marginalized posterior standard deviations ($\sigma$). The solid horizontal lines indicate the RMS distances averaged over all parameters in each panel. This plot shows nicely how different implementations of (even mildly) non-linear physics manifest themselves as larger biases in parameter estimation analyses. Moreover, although the RMS distance for {\HMCode} and {\Halofit} happen to be almost identical, the biases in each cosmological parameter can vary drastically (as e.g. in the case of $n_{\rm{s}}$).}
    \label{fig:code_effect}
\end{figure}

However, increasing the cut-off scale $k_{\rm max}$ to $0.4\hompc$ paints a different picture (see right panels of Figures \ref{fig:LCDM} and \ref{fig:wCDM_cc_0p4}): As on these scales the the power spectra predicted by the codes start to deviate more significantly from one another, in this case clear biases between the best-fit cosmologies computed with the three different predictors are seen. The values vary between 1 and $6\sigma$ depending on the parameter, but irrespective of the cosmological model. The averaged biases are about $0.5\sigma$ if a cut-off scale of $k_\mathrm{max}=0.2\hompc$ is employed and it is $1-2\sigma$ if $k_\mathrm{max}=0.4\hompc$. This aspect is visualized in a more condensed way in \autoref{fig:code_effect}. In this figure, the distances between the mean cosmology $\mu_{\rm{mod}}$ of each model to the fiducial cosmology is given in units of the 1D marginalized standard deviations $\sigma_{\rm{mod}}$ for each cosmological model, cut-off scale and cosmological parameter. The RMS distance (averaged over all cosmological parameters) is again indicated by the solid horizontal lines. It is most evident that increasing the cut-off scale from $0.2\hompc$ to $0.4\hompc$ has a major effect on this distance, both on average as well as for some of the individual cosmological parameters. This result is independent of the power spectrum predictor or cosmological model.

\begin{figure*}
  \centering
  \includegraphics[width=\textwidth]{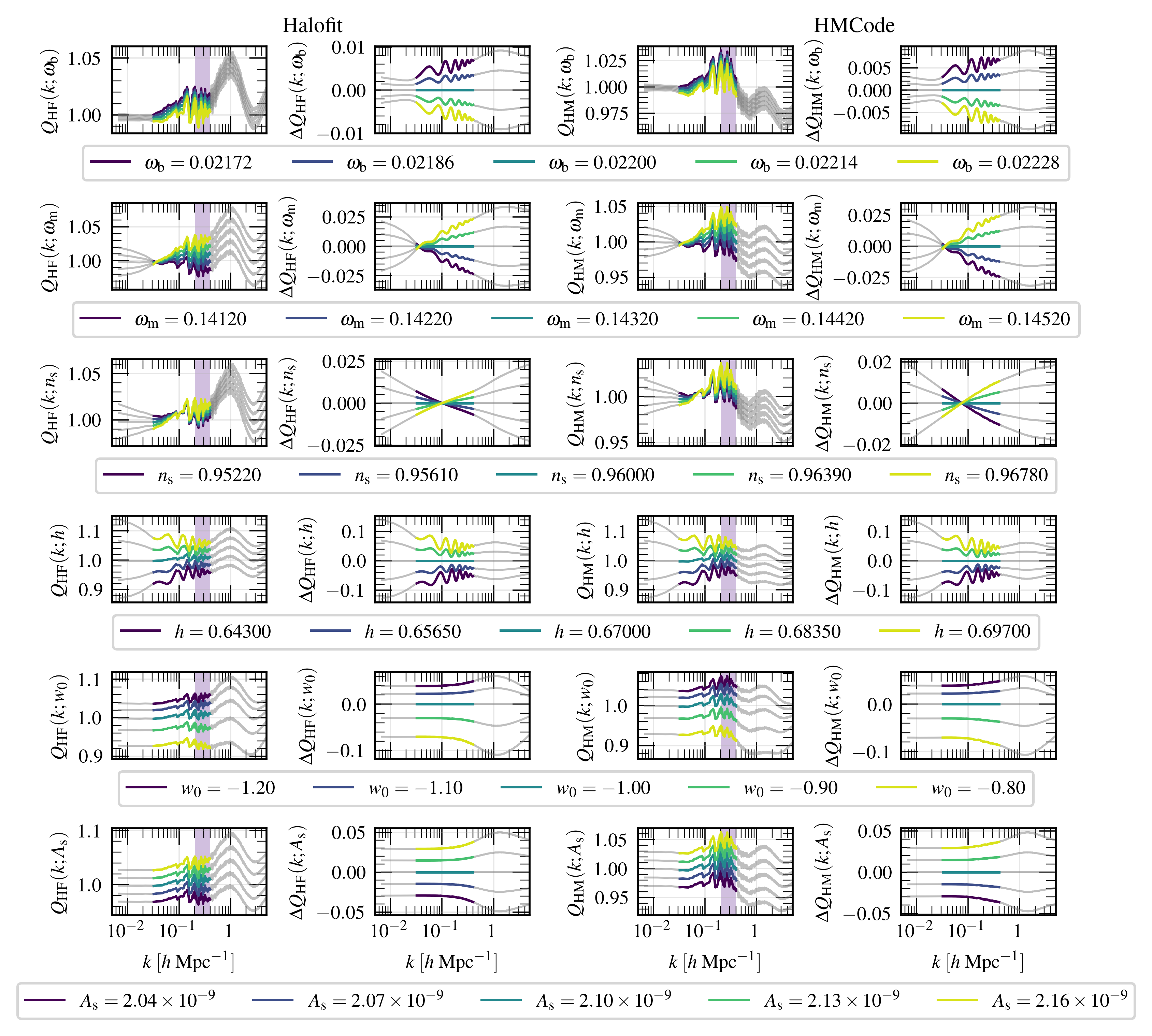}
  \caption{This figure shows the change of the power spectrum predicted by either {\Halofit} or {\HMCode} as individual cosmological parameters are varied while all other parameters are fixed to the fiducial cosmology. The comparison is shown in terms of the quantity $Q_{\rm{model}}$, being defined in the main text as the ratio of a power spectrum predicted by ${\rm{model}}\in\{{\rm{Halofit}},{\rm{HMCode}}\}$ with respect to the power spectrum produced by {\EEone} at the fiducial cosmology. The curves are shown plotted in the full range $k\in[0.005,5]$, however, only the colored parts are actually taken into account in the {\MCMCs} and hence only those $k$ modes have an effect on the biases observed in the {\MCMC} posteriors in this section. The shaded indigo bands mark the region between $k=0.2\hompc$ and $k=0.4\hompc$ which is responsible for the larger biases observed in the analysis with the larger cut-off scale. Notice the differences and similarities between this figure and \autoref{fig:var}.}
  \label{fig:var_wrt_fid}
\end{figure*}

We trace back the significant biases observed in the right panel plots in figures \ref{fig:LCDM} and \ref{fig:wCDM_cc_0p4} to the differences in the power spectrum predictions between the different predictor codes. To do so we first look at the quantities 
\begin{equation}
 Q_{\rm{mod}}(k; p) \equiv \frac{P_{\rm{mod}}(k;p)}{P_{\rm{EE}}(k;p_{\rm{fid}})}\,, 
\end{equation}
and
\begin{equation}
 \Delta Q_{\rm{mod}}(k; p) \equiv Q_{\rm{mod}}(k; p)-Q_{\rm{mod}}(k; p_{\rm{fid}})\,, 
\end{equation}
shown in \autoref{fig:var_wrt_fid}. Here ${\rm{mod}}\in\{{\rm{HF}}, {\rm{HM}}\}$, the set of investigated cosmological parameters is $p\in\{\omega_{\rm{b}}, \omega_{\rm{m}}, n_{\rm{s}}, h, w_0, A_{\rm{s}}\}$, $p_{\rm{fid}}$ corresponds to the respective parameter set to its fiducial value as given in \autoref{tab:FidCosmo}. The redshift is kept fixed at $z=1$. The quantity $Q$ differs from quantity $R$ defined in \autoref{eq:Rmodel} by keeping the denominator fixed to the fiducial cosmology, while it is also varied in the quantity $R$.

\begin{figure*}
	\includegraphics[width=0.8\textwidth]{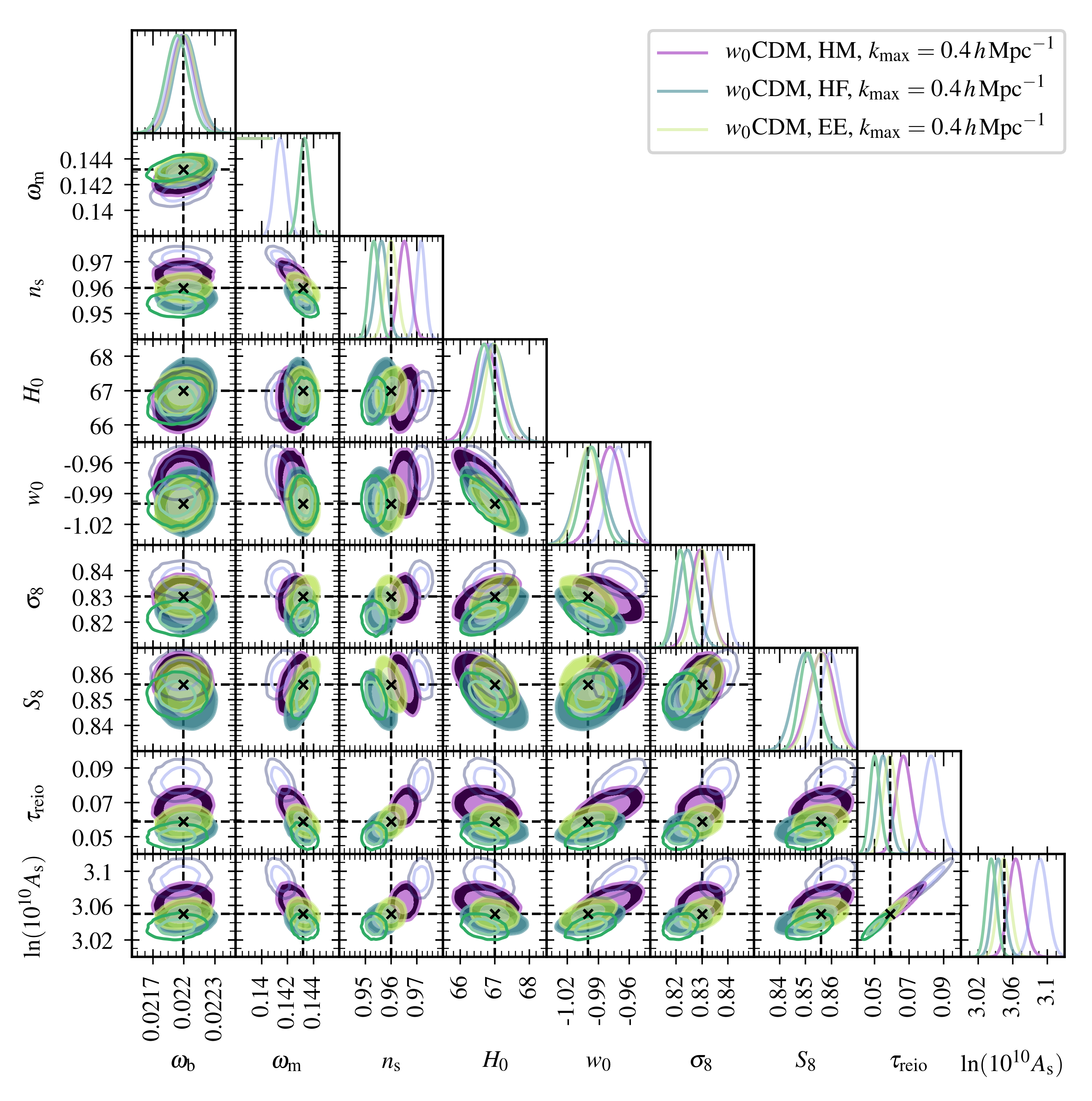}
	\caption{$w_0$CDM cross-comparison of {\HMCode} and {\Halofit} vs {\EEone} ($k_{\rm max} = 0.4 \hompc$) using the new uncertainty envelope $\sigma_{\rm{th,KB}}$ (filled contours) compared to the theoretical uncertainty envelope $\sigma_{\rm{th,S}}$ as given in \citet{Sprenger2019}. Notice that the empty contours correspond to the contours shown in \autoref{fig:wCDM_cc_0p4}. This plot nicely demonstrates how a carefully modelled theoretical uncertainty is capable of making the best-fit cosmology less model-dependent while keeping its credible contours reasonably tight. On the other hand, if a insufficient theoretical uncertainty is used, the resulting best-fit cosmology may end being significantly biased.}
	\label{fig:TUE_test}
\end{figure*}

Since the biases grow significantly as we increase the cut-off scale from $k=0.2\hompc$ to $k=0.4\hompc$, we shall focus primarily on this region in the following discussion (marked by the vertical indigo bands in \autoref{fig:var_wrt_fid}). The argument is most eye-catchingly illustrated considering the example of $n_{\rm{s}}$ (third row on \autoref{fig:var_wrt_fid}): while in general (i.e. over the entire $k$ range taken into account for the {\MCMCs}) the quantities $Q$ are very similar for {\Halofit} and for {\HMCode}, they show a difference in behaviour between $k=0.2\hompc$ and $k=0.4\hompc$. The curves turn upward for {\Halofit} but downward for {\HMCode}. In other words, at the small scale end of the considered $k$ range, {\Halofit} tends to overestimate the power spectrum while {\HMCode} underestimates it. The overestimation for {\Halofit} is however alleviated if the $n_{\rm{s}}$ value is decreased compared to the fiducial value. Similarly, the underestimation in the case of {\HMCode} is reduced if $n_{\rm{s}}$ is somewhat increased. Hence, the {\MCMC} favours lower $n_{\rm{s}}$ values for {\Halofit} and higher ones for {\HMCode} when fitting the fiducial cosmology. This aligns perfectly with what we observe in figures \ref{fig:LCDM} and \ref{fig:wCDM_cc_0p4}. The same argument can be applied to all other cosmological parameters, albeit with complex parameter degeneracies (made more complex by the inclusion of mock Planck data) resulting in shifts for most parameters.

E.g, changing $\omega_{\rm{m}}$ (second row on \autoref{fig:var_wrt_fid}) helps adjust for the differences on {\BAO} scales for {\HMCode} (note the posterior for $\omega_{\rm{m}}$ remains largely unchanged for {\Halofit} compared to {\EEone}), while also modulating the small scale amplitude shift from changing $n_{\rm{s}}$ with a similar but somewhat different scale dependence: where increasing $n_{\rm{s}}$ increases power on small scales and decreases power on large scales, instead decreasing $\omega_{\rm{m}}$ decreases power across the observable range, except for the very largest scales where the impact is negligible, with a larger effect on smaller scales.

Therefore, the combination of increased $n_{\rm{s}}$ and decreased $\omega_{\rm{m}}$ that we see for {\HMCode} results in a different impact on the matter power spectrum than changing $n_{\rm{s}}$ alone, giving a flatter increase in power from large to smaller scales.

Also, we see only a very small shift in the mean of each posterior for $H_0$, but instead see inflated posteriors. We can understand this as the mock Planck data not allowing for a considerably different mean value from the fiducial value, but that the sampler attempts to adjust for differences on {\BAO} scales by varying $H_0$, resulting in a decrease in sensitivity to $H_0$ compared to the {\EEone} case, where the non-linear prescription matches the fiducial model, due to washing out information from the placement of the {\BAO} peaks.

Finally, we note that $\omega_{\rm{b}}$ is largely unaffected by biases , as this parameter is very well determined by the mock Planck likelihood and cannot easily be shifted.

\paragraph*{Answer to question \#3:} While the choice of the power spectrum predictor only has a relatively minor impact on the best-fit cosmology and its uncertainties when mostly linear scales are considered, the impact grows as increasingly larger $k$ modes are taken into account for the fit. We found this result independently of the fact whether a simple $\Lambda$CDM or a more complex $w_0$CDM model was considered.

\subsection{Effect of theoretical uncertainty models on the posterior distribution}
\label{subsec:tue}

In the previous subsection, we have shown that the choice of the power spectrum predictor can, in some cases, have a quite significant effect on the resulting best-fit cosmology. These differences clearly cannot be physical but must be due to modelling and implementation differences. In order to account for this, as was discussed above, one has to include yet another source of uncertainty into the modelling: a theoretical uncertainty (see \autoref{sec:envelope}). However, a priori such an uncertainty is not known and hence it has to be modelled. For this reason it is a valid question to ask what the impact of the choice for a theoretical uncertainty model is. To answer this question we ran two sets of {\MCMCs} for {\EEone}, {\Halofit} and {\HMCode} (restricting ourselves to the $w_0$CDM model and the case where $k_{\rm max}=0.4\hompc$): one set with the theoretical uncertainty envelope as given in \citet{Sprenger2019} referred to as ``$\sigma_{\rm{th, S}}$'' and one set with the envelope described in \autoref{sec:envelope} referred to as ``$\sigma_{\rm{th, KB}}$''. More concretely, we first compute a fiducial galaxy power spectrum $P_{\rm{gg}}$ based on {\EEone}. In a second step, we run {\MCMCs} for all three predictors, always comparing to the same fiducial. This means that in the case of {\EEone} we perform an auto-comparison test, while for {\Halofit} and {\HMCode} cross-comparisons are performed.

The different choices of the theoretical uncertainty manifest themselves in different distances between the best-fit cosmologies and in different contours. Since $\sigma_{\rm{th, KB}}$ was modelled with specific information about the codes being used in this analysis while $\sigma_{\rm{th, S}}$ is a rather generic envelope, we expect the posterior distributions featuring $\sigma_{\rm{th, KB}}$ to agree significantly better with each other than the ones computed with $\sigma_{\rm{th, S}}$. Specifically, the more aggressive theoretical uncertainty $\sigma_{\rm{th, KB}}$ down-weights the non-linear portion and {\BAO} scales where the codes disagree relative to the linear portion of the galaxy power spectrum where the codes agree very well, so the bias is reduced. While this test mostly serves as a sanity check, it helps us understand the often neglected impact of modelling choices on the final cosmological parameter estimation.

At this point we shall reiterate: It is always possible to implement so much theoretical uncertainty into an {\MCMC} such that any two codes would agree with each other. Hence, while there is no real challenge in making the best-fit cosmologies of {\MCMCs} based on {\EEone}, {\Halofit} and {\HMCode} agree with each other, the difficulty lies in doing so without significantly inflating the credibility contours. 

\begin{figure}\includegraphics[width=\columnwidth]{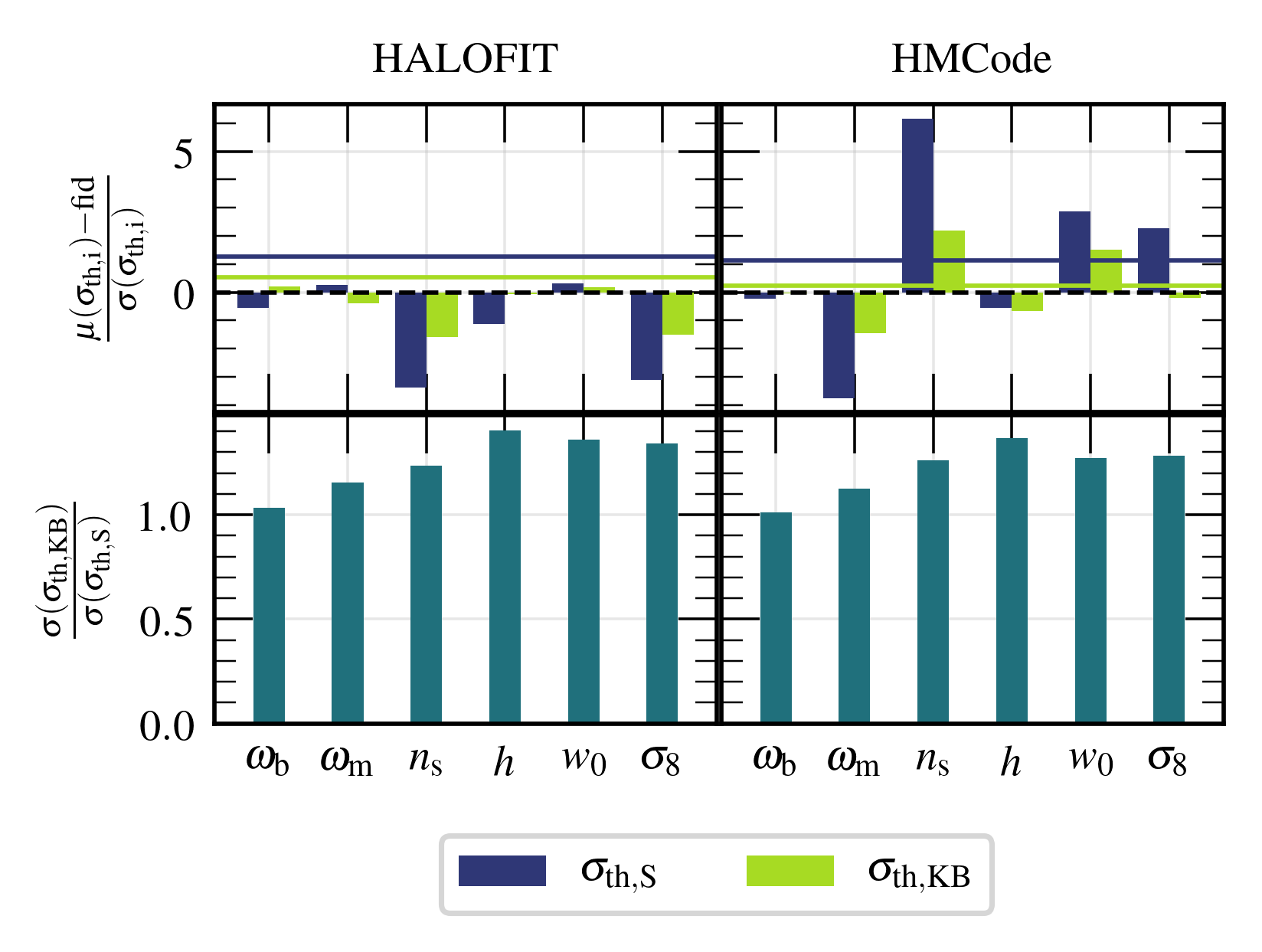}
	\caption{\textit{Upper row:} Differences between mean ($\mu$) and fiducial (fid) cosmology per cosmological parameter for {\Halofit} and {\HMCode} in the cross-comparison setup for both theoretical uncertainty envelopes $\sigma_{\rm th, i}$ tested in this paper: $i \in \{{\rm S}, {\rm KB}\}$. Here, only results for $w_0$CDM cosmologies with a cut-off scale of $k_{\rm max} = 0.4 \hompc$ are shown. The differences are again given in units of the respective 1D marginalized posterior standard deviations ($\sigma$). The blue bars in this plot correspond to the bars (both green and blue) in the lower right panel of \autoref{fig:code_effect}. Notice how $\sigma_{\rm{th, KB}}$ is able to reduce the RMS distance w.r.t. $\sigma_{\rm{th, S}}$ by a factor of $\sim2$ in the case of {\Halofit} and by a factor of $\sim5$ in the case of {\HMCode}. \textit{Bottom row:} Ratios between the $\sigma$'s measured with the two theoretical uncertainty envelopes. The fact that these ratios deviate only moderately from unity emphasizes the fact that the $\sigma_{\rm{th, KB}}$ is modeled carefully such that the agreement between the mean and the fiducial cosmology could be improved without drastically blowing up the uncertainty contours. The most extreme case is that for the parameter $h$ and {\Halofit}, where the theoretical uncertainty envelope causes $\sigma$ to increase by a factor of merely 1.4 compared to the contour computed with $\sigma_{\rm{th, S}}$. In all other cases $\sigma$ is increased by an even smaller factor.}
    \label{fig:tue_effect}
\end{figure}

Our test results are shown in  \autoref{fig:TUE_test} and more compactly in \autoref{fig:tue_effect} (for numerical results see \autoref{tab:newThU_crossconvres}). The filled contours in \ref{fig:TUE_test} represent the {\MCMCs} based on $\sigma_{\rm{th, KB}}$ and the empty contours those based on $\sigma_{\rm{th, S}}$. The empty contours correspond to the contours shown in \autoref{fig:wCDM_cc_0p4}. It is evident that the filled contours agree significantly better with each other than the empty ones while their areas are not much larger. The upper panels of \autoref{fig:tue_effect} show again the RMS distances (biases) between the mean cosmologies in units of the 68\% 1D marginalized posterior standard deviations $\sigma$ (similar to \autoref{fig:code_effect}). Clearly, the biases are smaller over all parameters when $\sigma_{\rm{th, KB}}$ is used compared to employing $\sigma_{\rm{th, S}}$. However, in order to paint the complete picture we need to take into account that not only the biases but also the uncertainties (i.e. the $\sigma$'s themselves) change when different uncertainty envelopes are used. The ratio between the standard deviations $\sigma$ depending on the uncertainty envelope $\sigma_{\rm{th},i},\;i\in\{{\rm{S}},{\rm{KB}} \}$ is shown in the lower panel. As expected, this ratio is always greater or equal to unity implying that the uncertainties are somewhat increased when $\sigma_{\rm{th, KB}}$ is used compared to using $\sigma_{\rm{th, S}}$. Note, however, how the maximal ratio is only $\sim 1.4$. This shows that a carefully modelled uncertainty envelope is capable of significantly reducing biases while not inflating the uncertainty contours by a lot.

\paragraph*{Answer to question \#4:} The choice of the theoretical uncertainty envelope $\sigma_{\rm th}$ can have a significant impact on the resulting estimate of the best-fit cosmological parameter values. A theoretical uncertainty which is not modelled with enough care may lead to a considerable bias in the best-fit cosmology. On the other hand, a carefully modelled theoretical uncertainty is able to account for model specifics and make the resulting posterior less dependent on the chosen power spectrum predictor while keeping the posterior credibility contours reasonably tight. We hence strongly suggest that in future cosmological parameter estimation studies more focus shall be put on the aspect of theoretical uncertainties in order to make the results more model-independent.

%% file: Chapters/06Conclusion.tex
\section{Conclusion}
\label{sec:conclusion}

In this paper, we have studied cosmological parameter inference results based on galaxy clustering observables in both {\LCDM} and {\wCDM} cosmologies focusing on the following four major questions:
\begin{enumerate}
    \item What are the performance differences of {\EEone}, {\Halofit} and {\HMCode}?
    \item What is the added value of considering (mildly) nonlinear scales in parameter forecasts?
    \item What is the impact of the choice of the (non-linear) predictor model on the parameter estimation result?
    \item How are the forecasting results affected by different choices of theoretical uncertainty models?
\end{enumerate}
To answer these questions, we have run multiple {\MCMCs} with the code {\MontePython}. The three different power spectrum predictors used for this work are {\EEone}, {\Halofit} and {\HMCode}. Notice that {\EEone} allows to predict the nonlinear correction only. We take advantage of this fact by modifying the Kaiser formula as described in appendix \ref{app:biasmodel}.

We start out our analysis (see \autoref{subsec:results_auto}) by a simple comparison of the {\MCMC} posteriors based on the three different power spectrum predictors. To do so, we let {\MontePython} estimate the parameters of a fiducial model, where both the fits and the fiducial are computed with the same predictor (referred to as ``auto-comparison'' tests). It is therefore expected that the best-fit cosmology recovers the fiducial cosmology almost perfectly in all cases. However, the posterior contours of {\EEone} tend to be tighter than the contours of {\Halofit} and {\HMCode}. The magnitude of this effect varies significantly depending on the cosmological model, the non-linear prescription and the $k$-mode cut-off employed. The minimal value is $2.2\%$ in the case of {\HMCode} vs. {\EEone} for {\LCDM} with $k_{\mathrm{max}}=0.2\hompc$ while the maximum value is $27.9\%$ the case of {\Halofit} vs. {\EEone} for {\wCDM} with $k_{\mathrm{max}}=0.4\hompc$. These results are shown in \autoref{fig:ac_sigma_analysis}. This observation together with the knowledge that the physics of non-linear clustering is more accurately modelled in {\EEone} than in the halo model-based models points at the fact that parameter forecasts based on {\EEone} are subject to smaller uncertainties than those based on the other two predictors.

We analyze the effect of including (mildly) non-linear scales on the posterior probability distribution of the {\MCMCs} in \autoref{subsec:kmax_uncertainties}. Concretely, we run two sets of {\MCMCs}, one considering only modes $k\leq k_{\rm max}=0.2\hompc$ (mostly linear scales) and one including also modes $k\leq k_{\rm max}=0.4\hompc$ (linear and mildly non-linear scales). We find that doing so reduces the 1D marginalized posterior standard deviations per parameter for all three different predictors (see \autoref{fig:kmax_effect}). This finding confirms that there is valuable information at smaller spatial scales that can be leveraged in order to learn more about the cosmological parameters (clearly this is the motivation for many modern cosmological surveys such as {\Euclid} to investigate data from those smaller scales). The magnitude of the effect varies between $\sim 9\%$ and $\sim 25\%$ (on average over all tested cosmological parameters) depending on the predictor and the cosmological model under consideration. Notice that we did not include scales above $0.4\hompc$ as we used only a simple linear galaxy bias model and we neglected baryonic effects in our analysis. The validity of such a simple model is known to break at even more non-linear scales. We thus consider the results presented in this paper as a proof of concept and leave a more sophisticated analysis of the impact of strongly non-linear scales on the posterior uncertainties to future work. Nevertheless, from these observations we can confirm that extending the range of considered $k$ modes beyond linear scales is certainly beneficial to a relevant degree for future cosmological parameter forecasts. In addition we find that, when more non-linear scales are considered in such an analysis, then the results depend on the prescription used to model the physics at these scales, unless a proper theoretical uncertainty is used.

In \autoref{sec:parameterestimation}, we go one step further and take on the view point of a mock data analysis than merely a parameter forecast. In a data analysis, we need to consider the possibility that the best-fit cosmology is biased compared to the true cosmology underlying the data. We approach the question about the bias caused by the non-linear model, which is employed in the parameter estimation analysis, by a set of {\MCMCs} referred to as ``cross-comparison tests'' (see \autoref{subsec:code_dependence}). For these tests, we always compute the fiducial power spectrum with {\EEone}. This fiducial power spectrum then serves as fake data which we try to fit with {\Halofit} and {\HMCode}. Doing so we make the following interesting finding (\autoref{fig:code_effect}): as long as only linear scales are considered, the biases between the best-fit cosmologies from all different codes are rather small (within the $1\sigma$ credible region). However, as soon as smaller scales are included in the {\MCMCs} the biases grow significantly, up to $6\sigma$. The take-away message from this finding is that the best-fit cosmology found in a parameter estimation may severely depend on the choice of the power spectrum predictor and hence the best-fit cosmology depends on the non-linear model, especially when more and more non-linear scales are taken into account. A similar finding was presented by~\cite{Martinelli2021} for the complementary case of cosmic shear, cautioning that care has to be taken when it comes to non-linear predictors and modelling in future analyses of high resolution large-scale structure data.

The interpretation from the previous paragraph calls for a more model-independent approach in parameter inference initiatives. We propose that this can be achieved by including not just an observational but also a carefully modelled theoretical uncertainty into the parameter estimation (the use of theoretical uncertainties in the field of parameter sensitivity forecasts is already known state-of-the-art). We test two different theoretical uncertainty envelopes in \autoref{subsec:tue}: a rather generic one and a carefully modelled one which is actually informed by the agreement of the three power spectrum predictors compared with each other. We find that the latter envelope leads to much smaller biases in the best-fit cosmologies while only marginally inflating the credible contours. By using our new theoretical uncertainty $\sigma_{\mathrm{th, KB}}$, the RMS distance between the mean and the fiducial cosmology can be reduced by a factor of $\sim2$ for {\Halofit} and even by a factor of $\sim5$ for {\HMCode}. This result is visualized in figures \ref{fig:tue_effect} and \ref{fig:TUE_test}. We thus advocate for the consideration of carefully modelled theoretical uncertainty envelopes in future data analyses as we demonstrate that they render the results much less model-dependent.

We conclude by stating that we have found the tightest posterior with the smallest bias between the best-fit and the fiducial cosmology by taking into account not just linear but also mildly non-linear scales and by including a well-informed theoretical uncertainty envelope into our model.

%% file: Apps/A_ImplementationDetails.tex
\section{Quasi-non-linear galaxy-galaxy power spectrum}
\label{app:biasmodel}
 
In \autoref{sec:analysis} of the main text we introduced how the {\NLC} computed by {\EEone} can be used to derive the observable galaxy-galaxy power spectrum. However, we glossed over a the fact that the relation we used is not fully self-consistent. This aspect is analyzed in more depth in this appendix.

\subsection{Review of linear {\RSDs} modelling}
\autoref{eq:PggExpansion} explains how the dark matter power spectrum $P_{\delta\delta}(k,z)$ is related to the observable galaxy-galaxy power spectrum $P_{\mathrm{gg}(k,\mu,z)}$. It is repeated here for convenience:
\begin{equation}
\begin{split}
    P_{\rm gg}(k,\mu,z) =&\quad f_{\rm AP}(z)\times f_{\rm res}(k,\mu,z)\times f_{\rm RSD}(\hat{k}, \hat{\mu}, z)\times b^2(z)\\
    &\times P_{\delta\delta}(k,z)\,,
\end{split}
\label{eq:PggExpansion_app}
\end{equation}
Further, in the same section we showed how the the {\NLC} $B$ computed by {\EEone} can be incorporated into that equation. In this section, we discuss the relation between $P_{\mathrm{gg}}$ and $B(k,z)$ in more detail, thereby focusing on the {\RSDs} ($f_{\mathrm{RSD}}$) and ignore the galaxy bias ($b$), Alcock-Paczynski ($f_{\rm AP}$), and resolution ($f_{\rm res}$) correction factors.

The observable galaxy-galaxy power spectrum depends on the direction of observation (see \autoref{eq:PggExpansion_app}) because in addition to the redshift signal due to the Hubble expansion there is a contribution to the redshift distance caused by the peculiar velocities of the galaxies, i.e.
\begin{equation}
    s = cz = H_0d+v_r
\end{equation}
where $s$ denotes the redshift distance to a galaxy, $c$ is the speed of light, $z$ is the measured redshift of a galaxy, $H_0$ the Hubble parameter at present day, $d$ is the proper distance to the galaxy and $v_r$ is the peculiar radial velocity. On large, linear scales the {\RSD} effect is described by the Kaiser formula \citep{Kaiser1987ClusteringSpace}, which relates the galaxy power spectrum $P_{\mathrm{gg}}$ to the dark matter power spectrum $P_{\delta\delta}$, the dark matter velocity power spectrum $P_{\theta\theta}$ ($\theta$ denoting the velocity divergence) and the dark matter mass-velocity cross-power spectrum $P_{\delta\theta}$:
\begin{equation}
\label{eq:nonlinearKaiser}
    P_{\rm gg}(k,\mu,z) = P_{\delta\delta}(k,z)+2\mu^2\beta P_{\delta\theta}(k,z)+\mu^4\beta^2P_{\theta\theta}(k,z)\,,
\end{equation}
where $\beta:=f/b(z)$ is the ratio of the growth rate $f$ and the galaxy bias $b$. Although \autoref{eq:nonlinearKaiser} still holds in the non-linear regime, in practice the computation of $P_\mathrm{gg}^\mathrm{non-linear}$ with this equation requires some means to compute $P_{\delta\delta}^\mathrm{non-linear}, P_{\delta\theta}^\mathrm{non-linear}$ and $P_{\theta\theta}^\mathrm{non-linear}$. Unfortunately, {\EEone}, {\Halofit} and {\HMCode} do only provide the capability to compute $P_{\delta\delta}^\mathrm{non-linear}$. However, to linear order the divergence of the velocity field is equal to the mass density field:
\begin{equation}
\label{eq:thetadelta}
    \theta(k) \approx \theta_\mathrm{lin}(k) = \frac{ik\delta(x)}{aHf} = \delta(k)
\end{equation}
such that $P_{\theta\theta}^{\rm linear}=P_{\delta\theta}^{\rm linear}=P_{\delta\delta}^{\rm linear}$. It is thus possible to simplify \autoref{eq:nonlinearKaiser} to
\begin{equation}
\begin{split}
\label{eq:linearKaiser}
    P_{\rm gg}^{\rm linear}(k,\mu,z) =& &&P_{\delta\delta}^{\rm linear}(k,z)\\
    &+&&2\mu^2fP_{\delta\delta}^{\rm linear}(k,z)\\
    &+&&\mu^4f^2P_{\delta\delta}^{\rm linear}(k,z)\\
    =&&& \left[1+\mu^2f\right]^2P_{\delta\delta}^{\rm linear}(k,z)
\end{split}
\end{equation}
On smaller scales, the Fingers of God effect \citep{1978IAUS...79...31T,Jackson:2008yv} adds to the {\RSD} and is accounted for by an additional prefactor \citep{Bull:2014rha} such that in total
\begin{equation}
\label{eq:simplelinearKaiser}
    P_{\rm gg}^{\rm linear}(k,\mu,z) = f_\mathrm{RSD}P_{\delta\delta}^{\rm linear}(k,z)
\end{equation}
with
\begin{equation}
    \label{eq:fRSD}
    f_\mathrm{RSD} = \left[1+\mu^2f\right]^2\exp{\left(-\hat{k}^2\hat{\mu}^2\sigma_{\mathrm{NL}}^2\right)}.
\end{equation}

\subsection{Quasi-non-linear extension of the {\RSDs}}
In the previous subsection we summarized the well-known Kaiser formula that describes the {\RSDs} of the linear power spectrum. Because we are interested in analyses including (mildly) non-linear scales in this paper, a generalization of that Kaiser formula to the non-linear galaxy-galaxy power spectrum is required. The obvious generalization is (compare to \autoref{eq:simplelinearKaiser})
\begin{equation}
    \label{eq:qnlKaiser}
    P_{\rm gg}^{\rm QNL1}(k,\mu,z) = f_\mathrm{RSD}P_{\delta\delta}^{\rm non-linear}(k,z)\,,
\end{equation}
where the superscript ``QNL'' stands for ``quasi-non-linear''. Notice however that this generalization breaks self-consistency as the relation \autoref{eq:thetadelta} cannot be generalized to the non-linear power spectrum. Let's expand the non-linear matter power spectrum into its linear contribution and the {\NLC}:
\begin{equation}
    \label{eq:qnl1}
    P_{\rm gg}^{\rm QNL1}(k,\mu,z) = f_\mathrm{RSD}P_{\delta\delta}^{\rm linear}(k,z)B(k,z)\,.
\end{equation}
Here, we started with \autoref{eq:nonlinearKaiser}, linearized this equation using \autoref{eq:thetadelta}, factored out $P_{\delta\delta}$ in \autoref{eq:linearKaiser} and replaced it by its non-linear counterpart. We shall refer to this as the ``QNL1'' model.

A second, self-consistent option to generalize \autoref{eq:simplelinearKaiser} is to keep the linear versions of the second and third term in \autoref{eq:nonlinearKaiser}, i.e. keep $P_{\delta\theta}^{\rm linear}=P_{\theta\theta}^{\rm linear}=P_{\delta\delta}^{\rm linear}$ and only use the non-linear version of $P_{\delta\delta}$ in the first term of \autoref{eq:nonlinearKaiser}. This model, referred to as ``QNL2'', is thus given by:
\begin{equation}
    \begin{split}
    \label{eq:qnl2}
    &P_{\rm gg}^{\rm QNL2}(k,\mu,z) \\
    =&P_{\delta\delta}^{\rm non-linear}(k,z)+2\mu^2fP_{\delta\theta}^{\rm linear}(k,z)+\mu^4f^2P_{\theta\theta}^{\rm linear}(k,z)\\
    =&P_{\delta\delta}^{\rm non-linear}(k,z)+2\mu^2fP_{\delta\delta}^{\rm linear}(k,z)+\mu^4f^2P_{\delta\delta}^{\rm linear}(k,z)
\end{split}
\end{equation}

Notice that self-consistency is not broken at the price of ignoring the non-linear contributions in $P_{\delta\theta}$ and $P_{\theta\theta}$ . If we substitute in the {\NLC} the linear matter power spectrum can be factored out again:
\begin{equation}
    P_{\rm gg}^{\rm QNL2}(k,\mu,z) = \left[B(k,z)+2\mu^2f+\mu^4f^2\right]P_{\delta\delta}^{\rm linear}(k,z)
\end{equation}
Adding the Fingers of God for {\RSDs} at small scales we arrive at 
\begin{equation}
    P_{\rm gg}^{\rm QNL2}(k,\mu,z) = f_\mathrm{RSD}^{\rm QNL2} P_{\delta\delta}^{\rm linear}(k,z)
\end{equation}
with the following expression for the {\RSD} correction factor:
\begin{equation}
    f_\mathrm{RSD}^{\rm QNL2} = \left[B(k,z)+2\mu^2f+\mu^4f^2\right]\exp{\left(-\hat{k}^2\hat{\mu}^2\sigma_{\mathrm{NL}}^2\right)}
\end{equation}

\begin{figure*}
	\includegraphics[width=0.495\textwidth]{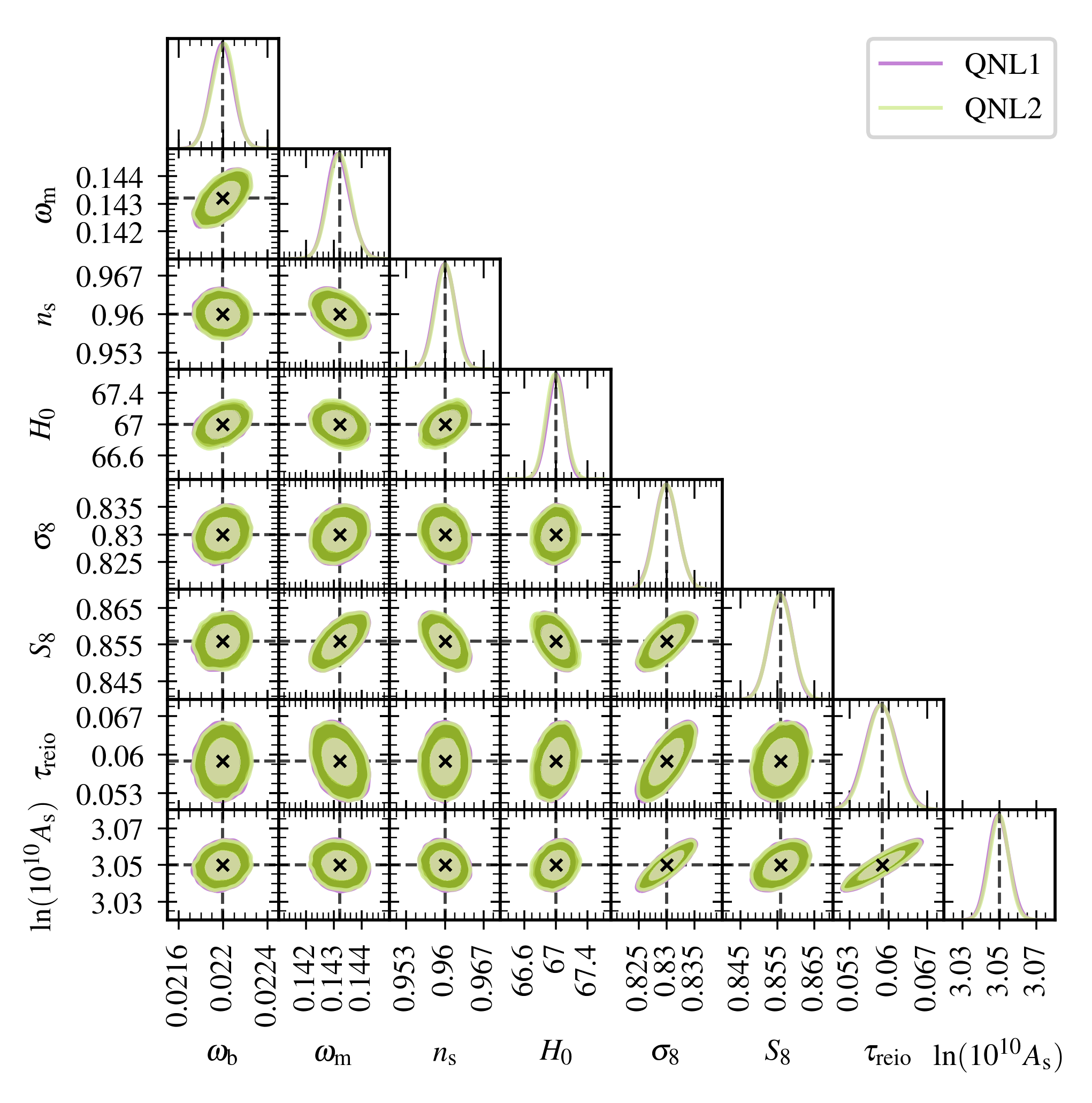}
	\includegraphics[width=0.495\textwidth]{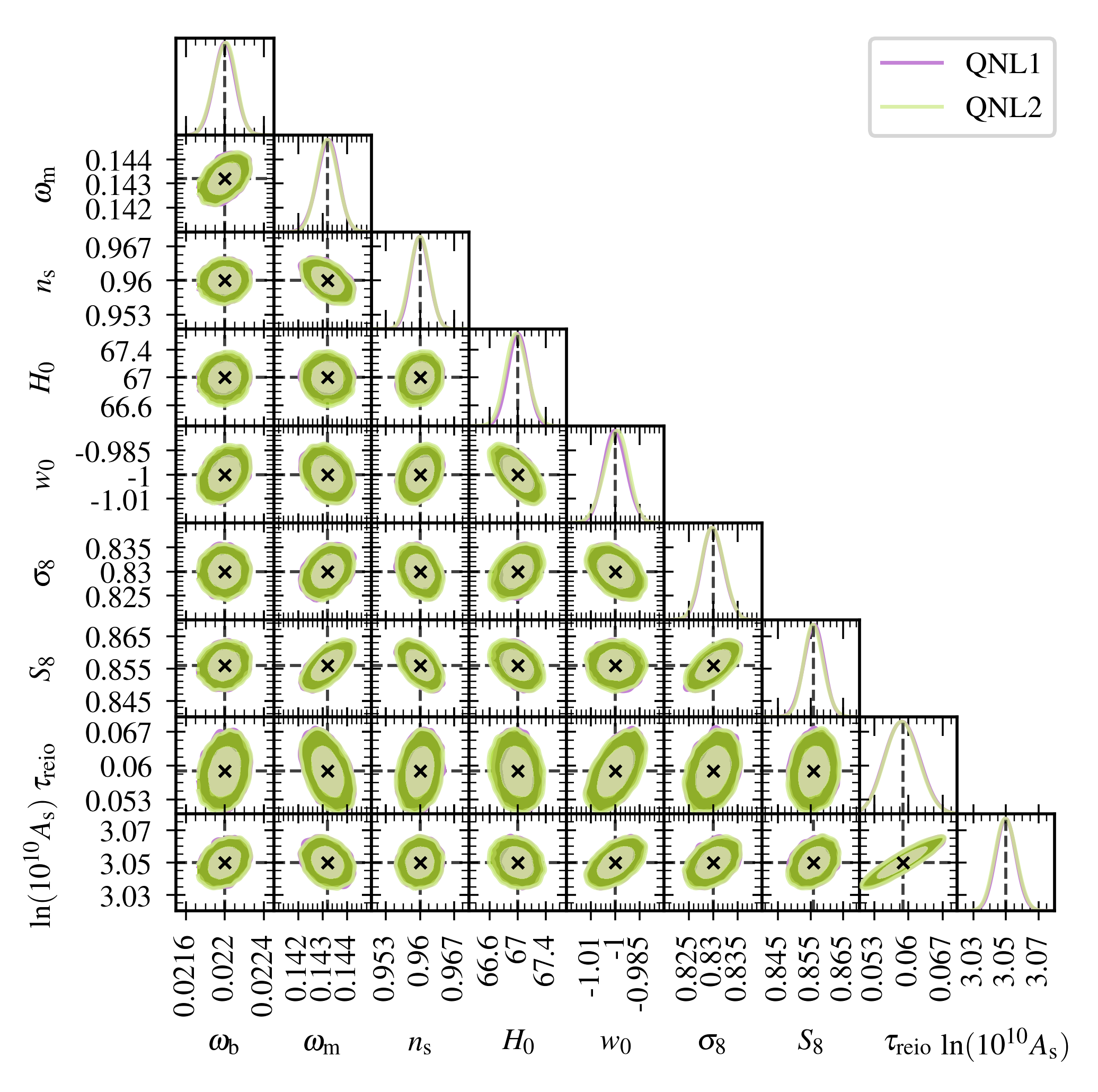}
	\caption{QNL1 vs QNL2 for {\LCDM} (top panel) and {\wCDM} (bottom panel). From these plots it becomes obvious that the two different QNL models (see \autoref{eq: nonlinear_kaiser1} and \ref{eq: nonlinear_kaiser2}) do only very marginally affect the posterior contours. The differences are so small that we neglect them throughout this entire paper.}
	\label{fig:wCDM_QNLcomp}
\end{figure*}

Having two options to generalize the Kaiser formula, the natural question arises by how much $P_{\rm gg}^{\rm QNL1}$ and $P_{\rm gg}^{\rm QNL2}$ differ from one another. Results of a test investigating this questions are shown in \autoref{fig:wCDM_QNLcomp}. As can be seen from these results, for the purpose of parameter forecasting there is virtually no difference between the QNL1 and QNL2 model. For this reason we treat the two models as one for the purposes described in the remaining analyses. More concretely, we have used model QNL2 in combination with {\EEone} because it is slightly more self-consistent than model QNL1. The latter, however, is simpler to implement when using {\Halofit} or {\HMCode}. Whenever we worked with the latter two codes, we hence applied QNL1, including for the fiducial model in the cross-comparison results of~\autoref{sec:parameterestimation}.

%% file: Apps/B_MCMCresults.tex
\section{MCMC results}
In this appendix we provide the numerical results of the {\MCMCs} (see tables) as well as the contours plots of the posterior probability distributions of the cosmological parameters as obtained in auto-comparison experiments (i.e. the sensitivity forecasting-like setup discussed in \autoref{sec:forecasts}).
\subsection{Auto-comparison results}
\begin{figure*}
    \includegraphics[width=0.495\textwidth]{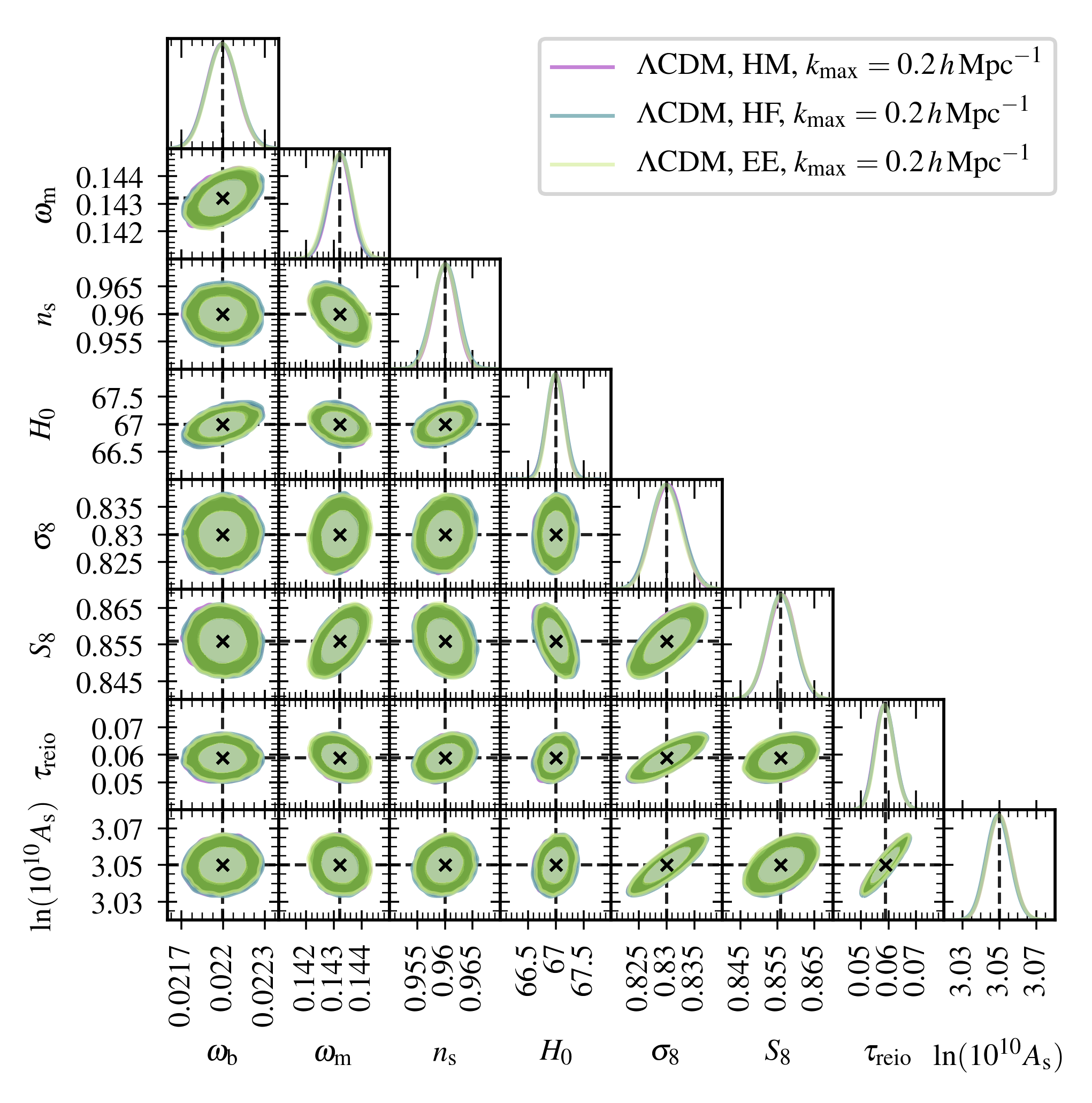}
    \includegraphics[width=0.495\textwidth]{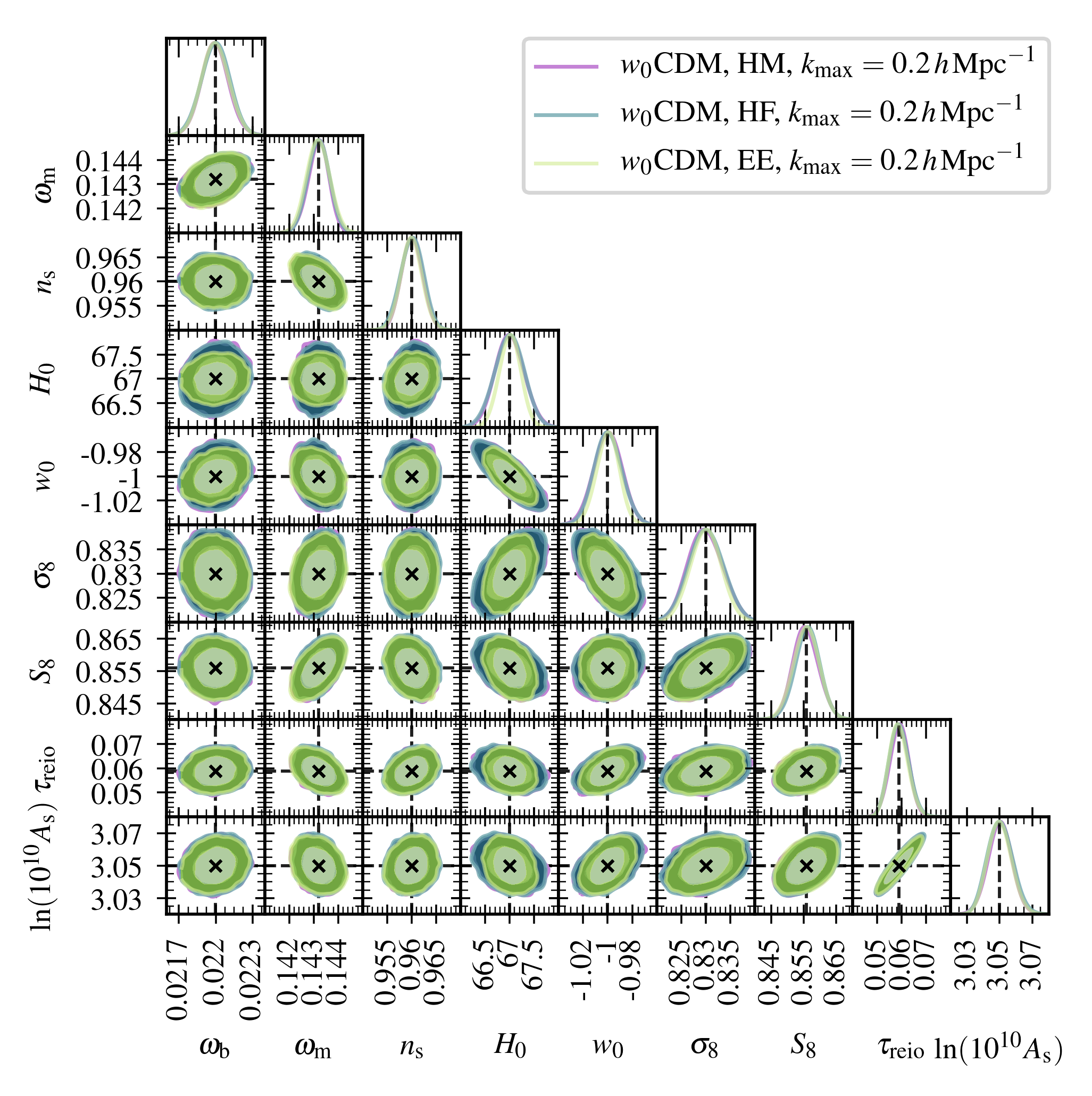}
	\caption{Auto-comparisons for {\LCDM} (top) and {\wCDM} (bottom) with cut-off scale $k_\mathrm{max}=0.2\hompc$. }
	\label{fig:ac_0p2}
\end{figure*}

\begin{figure*}
    \includegraphics[width=0.495\textwidth]{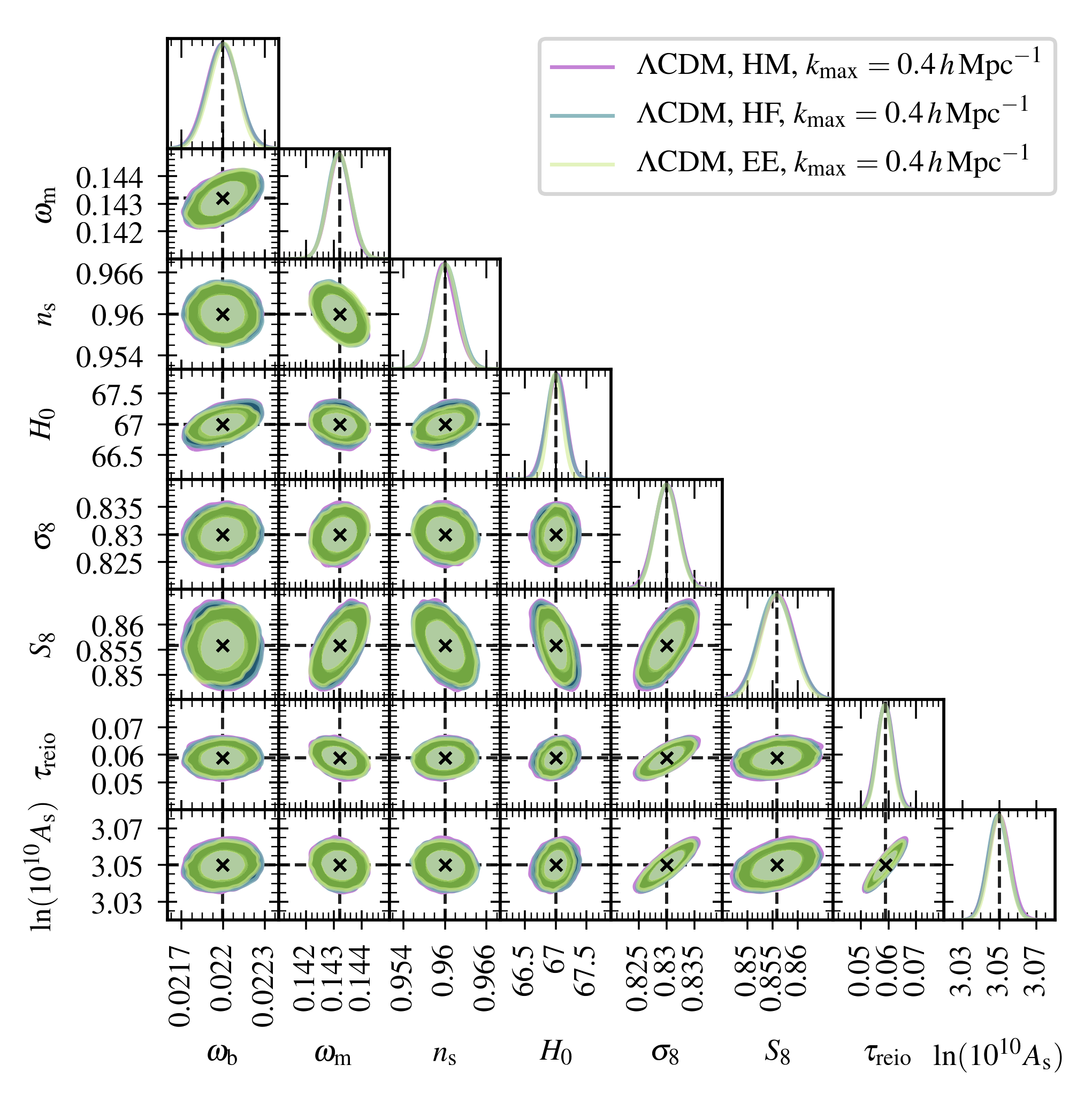}
    \includegraphics[width=0.495\textwidth]{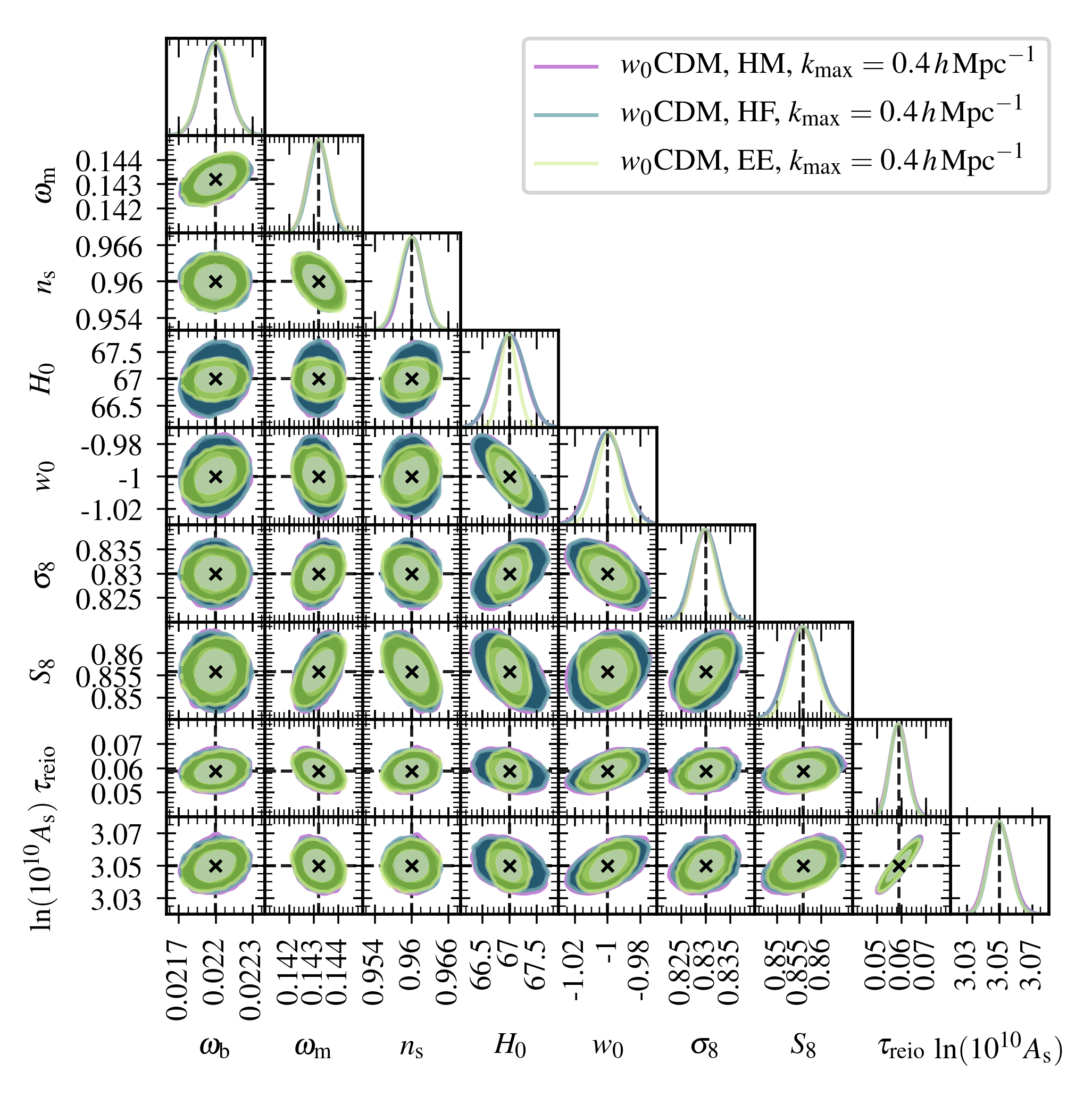}
	\caption{Auto-comparisons for {\LCDM} (top) and {\wCDM} (bottom) with cut-off scale $k_\mathrm{max}=0.4\hompc$. }
	\label{fig:ac_0p4}
\end{figure*}

In \autoref{sec:forecasts} we described the impact of the choice of the power spectrum predictor on forecasting results. As explained, in this setting only the various sensitivities are of interest as the mean posterior cosmologies coincide with the fiducial cosmology almost perfectly by construction. In the main text we provided the sensitivity information in the form of bar plots (see \autoref{fig:ac_sigma_analysis} and \ref{fig:kmax_effect}). As this representation may be somewhat unusual, we provide the more common contour plots here in this appendix (\autoref{fig:ac_0p2} and \ref{fig:ac_0p4}).

\begin{table*}
\centering%
\caption{Table of 1D marginalized posterior standard deviations ($\sigma$) for auto-comparisons. Each column reports about a power spectrum predictor that was used to both create the fiducial model and to sample the posterior probability surface. In each row of the table the standard deviation of such a posterior along a specific cosmological parameter axis and for a specific cosmological model and cut-off scale is denoted. Comparing the different predictors with each other it turns out that the posterior standard deviations of {\EEone} are usually smaller than those of {\Halofit} or {\HMCode}. The exception is $\omega_{\rm m}$ (and in some cases $n_{\rm{s}}$) which is itself an input parameter to {\EEone} while in {\Halofit} and {\HMCode} it is derived from $\omega_{\rm b}$ and $\omega_{\rm cdm}$. While in general the size of the posterior contours seem to be comparable for simple $\Lambda$CDM cosmologies and low cut-off scales, it can be seen that the difference in $\sigma$ between {\EEone} and {\Halofit} and {\HMCode} is larger for the case of $w_0$CDM cosmologies and the larger cut-off scale. The relative difference is sometimes as large as 70\% along a single parameter as in the case of the Hubble parameter $h$ for $w_0$CDM and $k_{\rm max} = 0.4 \hompc$ (more clearly visible in the lower right panel of \autoref{fig:ac_sigma_analysis}). From this one can conclude that {\EEone} has a substantial positive effect on the confidence of a parameter estimation compared to the other two predictors.}
\label{tab:autoconvres}
\begin{tabular}{c|c|c|ccc}
\multicolumn{3}{c|}{} & EuclidEmulator (EE) & Halofit (HF) & HMCode (HM) \\
\hline
\hline
\multirow{10}{*}{$\Lambda$CDM}& \multirow{5}{*}{0.2 $\hompc$} & $\sigma(\omega_{\mathrm{b}})$ & 1.11e-04 & 1.16e-04 & 1.15e-04 \\
& & $\sigma(\omega_{\mathrm{m}})$ & 4.49e-04 & 4.22e-04 & 4.24e-04  \\
& & $\sigma(n_{\mathrm{s}})$ & 2.16e-03 & 2.34e-03 & 2.15e-03 \\
& & $\sigma(h)$ & 1.49e-03 & 1.64e-03 & 1.62e-03 \\
& & $\sigma(\sigma_8)$ & 2.79e-03 & 2.83e-03 & 2.74e-03 \\
\cline{2-6}
& \multirow{5}{*}{0.4 $\hompc$} & $\sigma(\omega_{\mathrm{b}})$ & 1.04e-04 & 1.16e-04 & 1.16e-04 \\
& & $\sigma(\omega_{\mathrm{m}})$ & 4.18e-04 & 4.09e-04 & 4.11e-04  \\
& & $\sigma(n_{\mathrm{s}})$ & 1.83e-03 & 1.89e-03 & 1.79e-03 \\
& & $\sigma(h)$ & 1.20e-03 & 1.53e-03 & 1.58e-03  \\
& & $\sigma(\sigma_8)$ & 2.07e-03 & 2.18e-03 & 2.28e-03   \\
\hline
\hline
\multirow{12}{*}{$w_0$CDM} & \multirow{6}{*}{0.2 $\hompc$} & $\sigma(\omega_{\mathrm{b}})$ & 1.12e-04 & 1.18e-04 & 1.15e-04  \\
& & $\sigma(\omega_{\mathrm{m}})$ & 4.74e-04 & 4.37e-04 & 4.48e-04 \\
& & $\sigma(n_{\mathrm{s}})$ & 2.26e-03 & 2.35e-03 & 2.15e-03  \\
& & $\sigma(h)$ & 2.24e-03 & 3.02e-03 & 3.02e-03  \\
& & $\sigma(w_0)$ & 9.36e-03 & 1.15e-02 & 1.17e-02 \\
& & $\sigma(\sigma_8)$ & 3.23e-03 & 3.68e-03 & 3.58e-03 \\
\cline{2-6}
& \multirow{6}{*}{0.4 $\hompc$} & $\sigma(\omega_{\mathrm{b}}$ & 1.11e-04 & 1.14e-04 & 1.17e-04  \\
& & $\sigma(\omega_{\mathrm{m}})$ & 4.37e-04 & 4.17e-04 & 4.26e-04  \\
& & $\sigma(n_{\mathrm{s}})$ & 1.95e-03 & 1.91e-03 & 1.82e-03  \\
& & $\sigma(h)$ & 1.62e-03 & 2.74e-03 & 2.77e-03  \\
& & $\sigma(w_0)$ & 7.56e-03 & 9.97e-03 & 1.02e-02  \\
& & $\sigma(\sigma_8)$ & 2.34e-03 & 2.74e-03 & 2.81e-03 \\
\hline
\hline
\end{tabular}
\end{table*}

\subsection{Cross-comparison results}
We have discussed the impact of the choice of the power spectrum predictor on the final parameter estimation results in \autoref{sec:parameterestimation}. There we have provided the graphical summaries of the ``cross-comparison'' tests. Here, however, we provide the underlying numerical results in tabular form in order to assert full transparency. \autoref{tab:crossconvres} lists the mean cosmology and the $68\%$ credibility limits per cosmological parameter and cosmological model for both power spectrum predictors, {\Halofit} and {\HMCode}, with the fiducial cosmology given by {\EEone}. For all results in that table the uncertainty envelope by \citet{Sprenger2019} was used. In \autoref{tab:newThU_crossconvres}, on the other hand, we show the same quantities for the test with our new uncertainty envelope employed, that we have shown as lime-green lines in \autoref{fig:theoruncertainty}. Notice that for this experiment the results shown for {\EEone} correspond to an auto-comparison test while those for {\Halofit} and for {\HMCode} are cross-comparison tests.
\begin{table*}
\centering
\caption{Mean posterior cosmologies and $68\%$ credible intervals for {\Halofit} and {\HMCode} as resulting from cross-comparison {\MCMCs} where {\EEone} data serves as mock data. The fiducial values for the cosmological parameters are given in the first data column. Notice that for all results presented in this table the theoretical uncertainty by \citet{Sprenger2019} is used (for results based on the new theoretical uncertainty introduced in this paper see \autoref{tab:newThU_crossconvres}).}
\label{tab:crossconvres}
\begin{tabular}{c|c|c|ccc}
\multicolumn{3}{c|}{} &  Fiducial & Halofit (HF) & HMCode (HM) \\
\hline
\hline
\multirow{10}{*}{$\Lambda$CDM}& \multirow{5}{*}{0.2 $\hompc$} & $\omega_{\mathrm{b}}$ & $0.02200$ &$0.02193\pm 0.00012$ & $0.02194\pm 0.00011$ \\
& & $\omega_{\mathrm{m}}$ & $0.1432$ & $0.1431\pm 0.00044$ & $0.1428\pm 0.00044$ \\
& & $n_{\mathrm{s}}$ & $0.9600$ & $0.9593\pm 0.00237$ & $0.9577\pm 0.00217$ \\
& & $h$ & $0.6700$ & $0.6696\pm 0.00170$ & $0.6705\pm 0.00168$ \\
& & $\sigma_8$ & $0.8300$ &$0.8287\pm 0.00278$ & $0.8278\pm 0.00270$ \\
\cline{2-6}
& \multirow{5}{*}{0.4 $\hompc$} & $\omega_{\mathrm{b}}$ & $0.0220$ & $0.02193\pm 0.00011$ & $0.02192\pm 0.00012$ \\
& & $\omega_{\mathrm{m}}$ & $0.1432$ & $0.1433\pm 0.00040$ & $0.1418\pm 0.00043$ \\
& & $n_{\mathrm{s}}$ & $0.9600$ & $0.9535\pm 0.00194$ & $0.9715\pm 0.00184$ \\
& & $h$ & $0.6700$ & $0.6676\pm 0.00158$ & $0.6753\pm 0.00161$ \\
& & $\sigma_8$ & $0.8300$ & $0.8223\pm 0.00202$ & $0.8405\pm 0.00252$ \\
\hline
\hline
\multirow{12}{*}{$w_0$CDM} & \multirow{6}{*}{0.2 $\hompc$} & $\omega_{\mathrm{b}}$ & $0.0220$ & $0.02194\pm 0.00012$ & $0.02193\pm 0.00012$ \\
& & $\omega_{\mathrm{m}}$ & $0.1432$ & $0.1430\pm 0.00044$ & $0.1429\pm 0.00045$ \\
& & $n_{\mathrm{s}}$ & $0.9600$ & $0.9593\pm 0.00239$ & $0.9576\pm 0.00213$ \\
& & $h$ & $0.6700$ & $0.6679\pm 0.00305$ & $0.6710\pm 0.00266$ \\
& & $w_0$ & $-1.0000$ & $-0.9926\pm 0.01146$ & $-1.0027\pm 0.01044$ \\
& & $\sigma_8$ & $0.8300$ &$0.8270\pm 0.00361$ & $0.8283\pm 0.00339$ \\
\cline{2-6}
& \multirow{6}{*}{0.4 $\hompc$} & $\omega_{\mathrm{b}}$ & $0.02200$ & $0.02193\pm 0.00011$ & $0.02197\pm 0.00012$ \\
& & $\omega_{\mathrm{m}}$ & $0.1432$ & $0.1433\pm 0.00042$ & $0.1415\pm 0.00046$ \\
& & $n_{\mathrm{s}}$ & $0.9600$ & $0.9535\pm 0.00194$ & $0.9715\pm 0.00186$ \\
& & $h$ & $0.6700$ & $0.6669\pm 0.00274$ & $0.6684\pm 0.00288$ \\
& & $w_0$ & $-1.0000$ & $-0.9969\pm 0.00974$ & $-0.9695\pm 0.01065$ \\
& & $\sigma_8$ & $0.8300$  &$0.8216\pm 0.00272$ & $0.8367\pm 0.00292$ \\
\hline
\hline
\end{tabular}
\end{table*}

\begin{table*}
\centering
\caption{Mean posterior cosmologies and $68\%$ credible intervals for {\Halofit} and {\HMCode} as resulting from cross-comparison {\MCMCs} where {\EEone} data serves as mock data. The fiducial values for the cosmological parameters are given in the first data column. All results presented in this table are based on the new theoretical uncertainty given in \autoref{subsubsec:KBUncertainty}.}
\label{tab:newThU_crossconvres}
\begin{tabular}{c|c|c|ccc}
\multicolumn{3}{c|}{} &  EuclidEmulator (EE) & Halofit (HF) & HMCode (HM) \\
\hline
\hline
\multirow{6}{*}{$w_0$CDM} &  \multirow{6}{*}{0.4 $\hompc$} & $\omega_{\mathrm{b}}$ & $0.02200\pm 0.00011$ & $0.02202\pm 0.00012$ & $0.02199\pm 0.00012$ \\
& & $\omega_{\mathrm{m}}$ & $0.1432\pm 0.00050$ & $0.1430\pm 0.00048$ & $0.1424\pm 0.00052$ \\
& & $n_{\mathrm{s}}$ & $0.9600\pm 0.00234$ & $0.9562\pm 0.00240$ & $0.9651\pm 0.00235$ \\
& & $h$ & $0.6700\pm 0.00280$ & $0.6697\pm 0.00385$ & $0.6674\pm 0.00394$ \\
& & $w_0$ & $-0.9997\pm 0.01090$ & $-0.9977\pm 0.01324$ & $-0.9794\pm 0.01355$ \\
& & $\sigma_8$ & $0.8300\pm 0.00328$  &$0.8246\pm 0.00365$ & $0.8293\pm 0.00375$ \\
\hline
\hline
\end{tabular}
\end{table*}